\documentclass{aa}
\usepackage{graphicx}
\usepackage{longtable}
\usepackage{natbib} 
\bibpunct{(}{)}{;}{a}{}{,}
\usepackage{txfonts}
\usepackage{siunitx}
\usepackage{enumitem}
\usepackage{float}
\usepackage{placeins}
\usepackage{mwe,tikz}
\usepackage[colorlinks=true,linkcolor=blue,citecolor=blue,urlcolor=blue]{hyperref}
 
 \def\Halpha{$\mathrm{H\alpha}$}

\newcommand{\msunpyr}{\ifmmode{\,M_{\odot}\,\mbox{yr}^{-1}} \else{ M$_{\odot}$/yr}\fi}

\newcommand{\kms}{\ifmmode{\,\mbox{km}\,\mbox{s}^{-1}}\else{km\,s^{-1}}\fi}
\newcommand{\kpc}{\ifmmode {\,\mbox{kpc}} \else{kpc}\fi}
\newcommand{\msun}{\ifmmode M_{\odot} \else M$_{\odot}$\fi}
\newcommand{\rsun}{\ifmmode R_{\odot} \else R$_{\odot}$\fi}
\newcommand{\lsun}{\ifmmode L_{\odot} \else L$_{\odot}$\fi}
\newcommand{\zsun}{\ifmmode Z_{\odot,\,\text{Fe-group}} \else $Z_{\odot\,\text{Fe}}$\fi}
\newcommand{\xsun}{\ifmmode X_{\odot} \else $X_{\odot}$\fi}
\newcommand{\velo}{\ifmmode\varv\else$\varv$\fi}
\newcommand{\vinf}{\ifmmode\velo_\infty\else$\velo_\infty$\fi}
\newcommand{\rgal}{\ifmmode \,R_{\mathrm{gal}} \else R$_{\mathrm{gal}}$\fi}

\begin{document}

 	\title{New empirical mass-loss recipe for UV radiation line-driven winds of hot stars across various metallicities}
 	
 	\subtitle{}
 	\titlerunning{New empirical mass-loss recipe for hot stars across various metallicities}
 	
 	\author{D.~Pauli\inst{\ref{inst1},\ref{inst2}} 
          \and L.\,M.~Oskinova\inst{\ref{inst1}}
          \and W.-R.~Hamann\inst{\ref{inst1}}
          \and A.\,A.\,C.~Sander\inst{\ref{inst3}} 
          \and Jorick\,S.\,Vink\inst{\ref{inst4}}
          \and M.~Bernini-Peron\inst{\ref{inst3}}
          \and J.~Josiek\inst{\ref{inst3}} 
          \and R.\,R.~Lefever\inst{\ref{inst3}}
          \and H.~Sana\inst{\ref{inst2}}
          \and V.~Ramachandran\inst{\ref{inst3}}
        }
 	
 	\authorrunning{D. Pauli et al.}
 	
 	\institute{
        Institut f{\"u}r Physik und Astronomie, Universit{\"a}t Potsdam, Karl-Liebknecht-Str. 24/25, 14476 Potsdam, Germany\label{inst1}
 	    \and Institute of Astronomy, KU Leuven, Celestijnenlaan 200D, 3001 Leuven, Belgium\label{inst2}
        \and Zentrum f{\"u}r Astronomie der Universit{\"a}t Heidelberg, Astronomisches Rechen-Institut, M{\"o}nchhofstr. 12-14, 69120 Heidelberg\label{inst3}
        \and Armagh Observatory and Planetarium, College Hill, BT61 9DG Armagh, Northern Ireland\label{inst4}
 	}
 	
 	\date{Received ; Accepted}
 	
 	\abstract
         {The winds of massive stars remove a significant fraction of their mass, strongly impacting their evolution. As a star evolves, the rate at which it loses mass changes. In stellar evolution codes, different mass-loss recipes are employed for different evolutionary stages. The choice of the recipes is user-dependent and the conditions for switching between them are poorly defined.
          }
        {Focusing on hot stars, we aim to produce a physically motivated, empirically calibrated mass-loss recipe suitable for a wide range of metallicities. We want to provide a ready-to-use universal recipe that eliminates the need for switching between recipes for hot stars during stellar evolution calculations.
        }
        {We compile a sample of hot stars with reliable stellar and wind parameters in the Galaxy and the Magellanic Clouds. Our sample spans effective temperatures from ${T\approx\SIrange{12}{100}{kK}}$ and initial masses from ${M_\mathrm{ini}\approx15\,\msun\text{\,--\,}150\,\msun}$. The sample is used to determine the dependence of the mass-loss rate on the basic stellar parameters.}
        {We find that independent of evolutionary stage and temperature, the wind mass-loss rate is a function of the electron-scattering Eddington parameter ($\Gamma_{\!\text{e}}$) and metallicity ($Z$), being in line with expectations of radiation-driven wind theory. Our derived scaling relation provides an adequate ($\Delta\log(\dot{M}/(\msunpyr))=0.43$) and broadly applicable mass-loss recipe for hot stars.
        }
        {The newly derived mass-loss recipe covers nearly the entire parameter space of hot stars with UV radiation-driven winds and eliminates the need for interpolation between mass-loss formulae at different evolutionary stages when applied in stellar evolution models. Examples of stellar evolution calculations using our new recipe reveal that the predictions on the ionizing fluxes and final fates of massive stars, especially at low metallicity, differ significantly from models that use the standard mass-loss rates, impacting our understanding of stellar populations at low metallicity and in the young Universe. 
        }
 	
 	\keywords{ stars: mass-loss -- stars: winds, outflows -- stars: atmospheres -- stars: massive -- stars: early-type -- stars: evolution
        }
 	\maketitle

 	\section{Introduction} 
 	\label{sec:intro} 

        Massive stars are the mighty engines of the Universe. They shape their surroundings through intense ionizing radiation and powerful stellar outflows, making them regulators of star formation and contributors to the chemical enrichment of galaxies \citep[e.g.,][]{mae1:83,dra1:03,hop1:14,ram1:18,cro1:19}. Despite their significance, understanding the evolutionary pathways of massive stars remains challenging due to uncertainties in nuclear reaction rates, internal mixing processes, and stellar winds. While a star’s initial mass is the primary factor determining its evolution, the effects of mass loss on the evolution -- particularly for the more massive ({$M_\mathrm{ini}>30\,\msun$}) stars -- cannot be neglected as stellar winds can remove significant portions of the hydrogen-rich envelope, altering the star’s evolutionary path and its final fate \citep[e.g.,][]{con1:76,mae1:87,lan1:87}.
        
        The theoretical foundation for understanding UV-radiation-driven stellar winds was established in the late 20th century \citep[e.g.,][]{mih1:72,mih1:76,cas1:75,abb1:82,pau1:86}. Over the past decades, significant advancements, both theoretical and empirical, have improved our understanding of the mass lost through stellar winds \citep[e.g.,][]{vin1:01,mok1:07,krt1:18,san2:20,ric1:22,bjo1:23}. However, these studies often focus on specific groups of stars, such as main-sequence stars, supergiants, or Wolf-Rayet (WR) stars, and sometimes only address a narrow range of metallicities. In stellar evolution models, this segmented approach is reflected in the use of different mass-loss prescriptions for distinct evolutionary phases, resulting in abrupt transitions or simplistic interpolations between these phases \citep[e.g.,][]{bro1:11,eks1:12,cho1:16,lim1:18}.

        When aiming to create synthetic stellar populations, the segmented approach of combining different mass-loss recipes introduces considerable uncertainty as several widely used mass-loss prescriptions overestimate mass-loss rates by at least an order of magnitude, particularly for low-metallicity stars \citep[e.g.,][]{osk1:07,kob1:19,she1:20}. This discrepancy leads to population synthesis models overpredicting intermediate-mass helium (He) and WR stars at low metallicities. Such inaccuracies have profound implications for our interpretation of spectra from distant, low-metallicity galaxies observed by the Hubble and James Webb Space Telescopes \citep{cur1:23,bun1:23}, as well as on our understanding of massive star feedback in galaxies. Furthermore, it impacts our understanding of the formation and evolution of compact objects, and our predictions for compact object mergers, which are now routinely observed by gravitational wave detectors like LIGO/VIRGO \citep[][ and references therein]{abb1:23}.

        In recent decades, extensive observing programs utilizing various ground-based and space telescopes have gathered multi-wavelength and multi-epoch spectroscopic data of hot stars in the Galaxy (GAL; $\zsun=0.014$; \citealt{mag1:22}) and in the low-metallicity environments of the Large Magellanic Cloud (LMC; $Z_\mathrm{LMC,\,\text{Fe-group}}=1/2\,\zsun$; \citealt{vin2:23}) and Small Magellanic Cloud (SMC; $Z_\mathrm{SMC,\,\text{Fe-group}}=1/7\,\zsun$; \citealt{vin2:23}). These data have been analyzed in several literature studies using state-of-the-art stellar atmosphere codes, providing stellar and wind parameters. By leveraging these comprehensive datasets, we aim to develop an updated mass-loss prescription, broadly applicable to hot stars with UV-driven winds and across a wide range of metallicities.

        Section~\ref{sec:Observations} of this paper describes the sample of hot stars used in this study and outlines the selection criteria applied. The resulting mass-loss prescription is presented in Sect.~\ref{sec:result}. A detailed comparison to commonly used mass-loss recipes and the implications of adopting our updated mass-loss prescription in stellar evolution models are discussed in Sect.~\ref{sec:discuss}. Finally, a summary of our findings is provided in Sect.~\ref{sec:conclusions}.
        
 	\section{Method}
 	\label{sec:Observations}
    \subsection{Parametrizing stellar mass-loss rates}

        \citet*[hereafter CAK]{cas1:75} established the first theoretical framework for radiatively driven stellar winds. In their zero-sound speed approximation, they can express the mass-loss rate as
        \begin{equation}
            \dot{M}\propto
            \alpha \left(1-\alpha\right)^{(1-\alpha)/\alpha} k^{1/\alpha}
            \Gamma_\text{\!e}^{1/\alpha}\left(1-\Gamma_\text{\!e}\right)^{-(1-\alpha)/\alpha},\label{eq:CAK}
        \end{equation}
        with $\Gamma_\text{\!e}=\kappa_\text{\!e} L/(4\pi c G M)$ being the electron scattering, or classical Eddington parameter. The quantities $k$ and $\alpha$ are so-called (line-)force multiplier parameters which have inherent temperature and metallicity dependencies \citep[see, e.g., table 3 in][]{pul1:00}.
        For most stars, the radiative force arising from line opacities is crucial to overcome gravity and launch the stellar wind, meaning that the line opacity cannot be ignored in a description of radiation-driven mass loss. While the parametric CAK description already simplifies the inherent complexity of the radiative force, the further dependencies of the force multiplier parameters prevent any straightforward analytic description\footnote{The famous ``cooking recipe'' in \citet{kud1:89} requires numerical iteration and thus is hardly used in current evolution modeling.}. Moreover, the validity of the CAK description is limited and for example, does not incorporate WR stars. Formally, the full complexity of the line force can be encapsulated in a $\Gamma_\text{rad}$ quantity including the full flux-weighted opacity, but while $\Gamma_\text{\!e}$ can be reasonably determined and is often approximately constant in the relevant part of the stellar atmosphere, $\Gamma_\text{\!rad}$ instead is a radially strongly varying function and thus not handy for any empirically calibrated prescription.
        
        Given these considerations, we choose a $\Gamma_\text{\!e}$-dependence for our $\dot{M}$-description. For stars that are not close to the Eddington limit (${\Gamma_\text{\!e}\ll1}$), the mass-loss rate's dependence on the Eddington factor can be further simplified to
        \begin{equation}
            \dot{M}\propto \Gamma_\text{\!e}^{n}.
        \end{equation}

        Within this approximation, the information on the metallicity dependence included in the full line opacities or the force multiplier parameters is lost. 
        To remedy this shortcoming, we use the well-established fact that hot star winds are driven by UV-radiation pressure on metal ions, in particular the line-rich ions of iron and the iron-group elements. Consequently, the mass-loss rate is expected to depend on the abundances of the iron group elements Sc, Ti, V, Cr, Mn, Fe, Co, and Ni ($Z_\text{Fe-group}$), meaning that $\dot{M}$ gets weaker as the metallicity decreases.
        
        As an approach for developing an empirical description, useful for implementation in stellar evolution calculations, we assume that the mass-loss rates of hot stars can be expressed by a power-law description depending only on $\Gamma_\text{\!e}$ and $Z_\text{Fe-group}$
        \begin{equation}
            \dot{M}\propto \Gamma_\text{\!e}^{n}\,Z_\text{Fe-group}^m.
        \end{equation}
        The metallicity exponent, $m$, is constrained by empirical and theoretical efforts to be between $m\approx \SIrange{0.5}{1.6}{}$ (see also Sect.~\ref{sec:discuss_z}).

        In many empirical studies of massive stars, mass-loss rates are often presented only as a function of luminosity, whereas stellar mass and surface H abundance are not considered. This simplification is only valid when comparing stars within specific evolutionary subgroups (e.g., main-sequence stars, supergiants) that have similar luminosity-to-mass ratios and surface hydrogen abundances. However, such relations can deviate by several orders of magnitude when applied to stars in different evolutionary states. 
  
    \subsection{Sample selection}
    \label{sec:sample}
        
     	\begin{figure*}[tbhp]
     	    \centering
     	    \includegraphics[width=\textwidth]{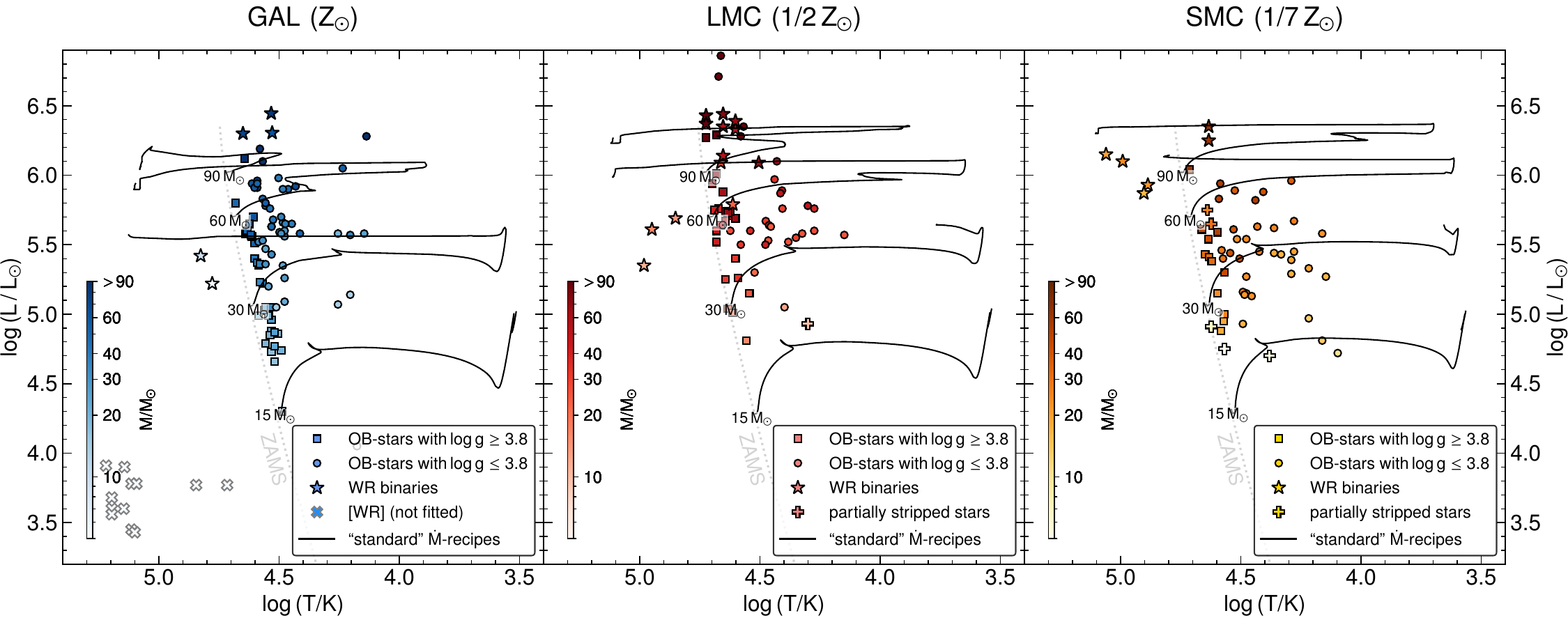}
     	    \caption{Hertzsprung-Russell diagrams (HRDs) containing our sample of the Galaxy (left), LMC (middle), and SMC (right). Overlaid are stellar evolution tracks computed with the standard wind mass-loss rates (see Sect.~\ref{sec:mesa} and Appendix~\ref{app:mesa}) and an initial rotational velocity of ${\varv_\mathrm{rot,ini}=100\,\mathrm{km\,s^{-1}}}$. The individual stars are color-coded based on their current spectroscopic mass, corrected for the effects of centrifugal force.}
     	    \label{fig:hrd_sample}
     	\end{figure*}
        
        Over the past decades, numerous studies have focused on the spectral analysis of hot, massive stars and the determination of their mass-loss rates. While we cannot rule out systematic biases, e.g., due to additional physics not covered in the current generation of atmosphere models, we assume within this work that the current spectroscopic analysis methods derive accurate measurements of the mass-loss rates within a reasonable uncertainty arising from approximations and assumptions made within the 1D stellar atmosphere codes. Recently, \cite{san1:24} demonstrated that the derived stellar parameters for the same star using different non-local thermodynamic equilibrium (non-LTE) stellar atmosphere codes are in good agreement. Therefore, to minimize systematic biases, we include in our sample only spectroscopic studies that use established non-LTE stellar atmosphere codes. Despite this, these studies cannot be used indiscriminately when aiming to derive a robust empirical mass-loss relation. To mitigate systematic differences, our selected sample adheres to the following criteria:        
        \begin{enumerate}
            \item A star must be sufficiently hot ($T_\text{eff}\gtrsim \SI{12}{kK}$) to be properly in the regime of radiation-driven winds.
            \item The distance to a star must be well-determined to ensure accurate luminosity estimates. This means that we require for bright Galactic targets that their distance is known via cluster associations, and for fainter Galactic targets their previous distance estimates have to agree with Gaia parallaxes.
            \item The analysis of a star must employ a non-LTE stellar atmosphere code that includes iron line blanketing (PoWR, CMFGEN, and Fastwind).
            \item Stars are excluded if their wind-line variability differs from the natural variability expected in OB stars, yielding differences in the mass-loss rates larger than $50\%$ \citep{mas1:24}.
            \item Stars suspected to belong to binary or multiple systems but analyzed as single stars are excluded.
            \item To avoid including potential, not yet discovered post-interaction binary stars, it is required that the spectroscopic mass (corrected for centrifugal force) of main-sequence stars must agree within a factor of $\sim$$1.5$ with the evolutionary mass. For stars lacking a spectroscopic mass estimate (e.g., WR stars), an orbital mass estimate is required for inclusion.
            \item The star must exhibit clear wind features in the UV or optical. For our sample, this corresponds to $\Gamma_{\!\text{e}}\gtrsim0.1$.
            \item For stars with optically thin winds (e.g., no or weak \Halpha\,\,emission), mass-loss rates must be derived from UV wind features. For stars with optically thick winds (e.g., WR stars or supergiants), mass-loss rates derived from optical spectra are accepted.
        \end{enumerate}

        It is well established that stellar winds are inhomogeneous. In stellar atmosphere codes, inhomogeneities are typically modeled using the ``microclumping'' approach, which assumes that the wind consists of optically thin clumps. The density enhancement of the clumps relative to a smooth wind with the same mass-loss rate can be characterized by a density contrast parameter $D$ \citep[e.g.,][]{hil1:91,ham1:98}. The resulting degeneracy of the clumping factor and the mass-loss rate can be broken by employing both UV and optical wind diagnostic lines. While in many spectral analyses included in our sample, the effect of microclumping is accounted for, some studies that only rely on optical spectra assume a smooth wind ($D=1$). However, numerous investigations have shown that the winds of OB-type stars are modeled the best with a clumping factor of $D=10$ \citep[e.g.,][]{sur1:13,ver1:24,san1:24}. Therefore, in cases where a mass-loss rate was provided for a smooth wind, the rates were corrected in a simple-minded way to the mass-loss rates of a star with a clumped wind assuming a clumping factor $D=10$, by using
        \begin{equation}
            \log(\dot{M}/(\msunpyr)) = \log(\dot{M}_\mathrm{smooth}/(\msunpyr))-\log(\sqrt{D}).
        \end{equation}
        Following a comprehensive literature review, our final sample comprises 70 stars from our Galaxy (GAL), 61 from the LMC, and 57 from the SMC. The sample includes OB-type main-sequence stars ($\log\,(g\,[\mathrm{cm\,s^{-2}}])>3.8$), OB-type giants and supergiants ($\log\,(g\,[\mathrm{cm\,s^{-2}}])<3.8$), Wolf-Rayet stars, and recently identified partially stripped stars with robust mass-loss estimates. As such, the sample spans a wide temperature range (${T\approx\SIrange{12}{100}{kK}}$) and initial masses (${M_\mathrm{ini}\approx15\,\msun\text{\,--\,}150\,\msun}$) (see Fig.\ref{fig:hrd_sample}). Detailed stellar and wind parameters for the sample, along with references, are provided in Tables~\ref{tab:stellar_parameters_summary_supergiants} to \ref{tab:stellar_parameters_summary_main_sequence}.
      
    \section{Results}
    \label{sec:result}

    \subsection{The strong winds of post-interaction binaries}
    
     	\begin{figure*}[tbhp]
     	    \centering
     	    \includegraphics[width=0.45\textwidth]{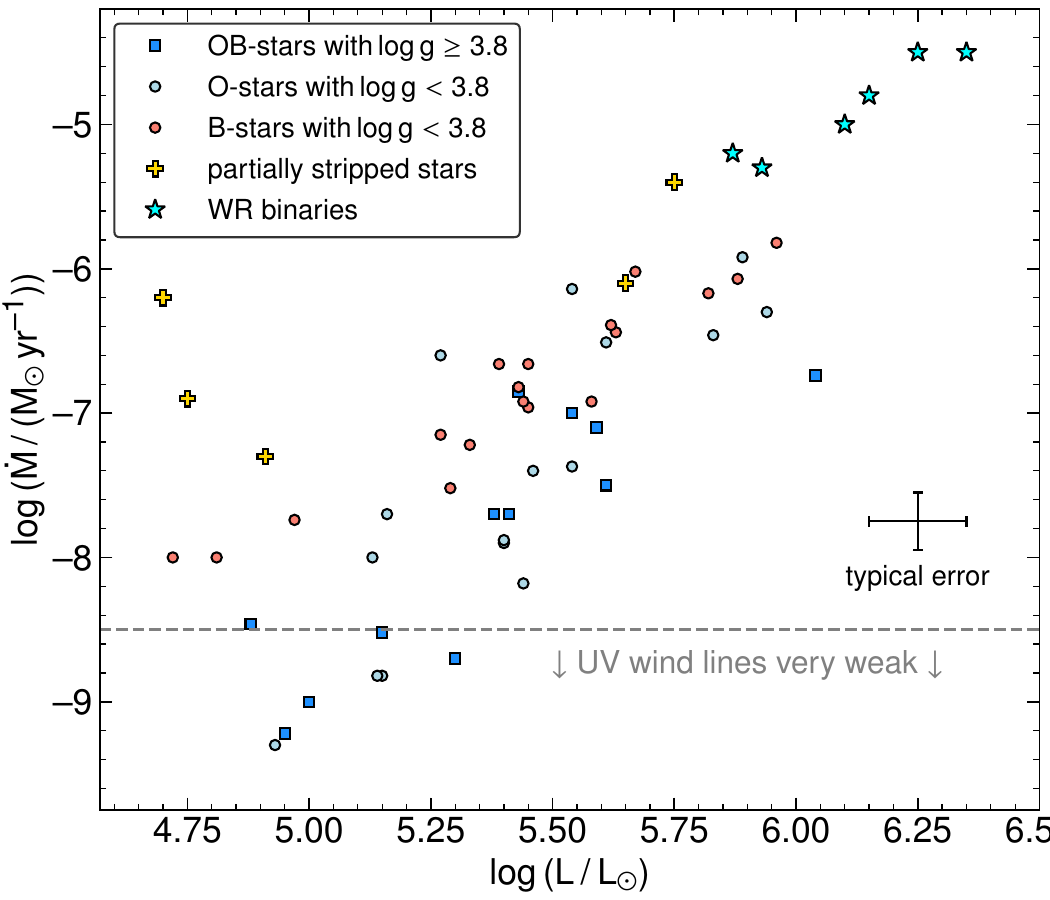}
            \hspace{3ex}
     	    \includegraphics[width=0.45\textwidth]{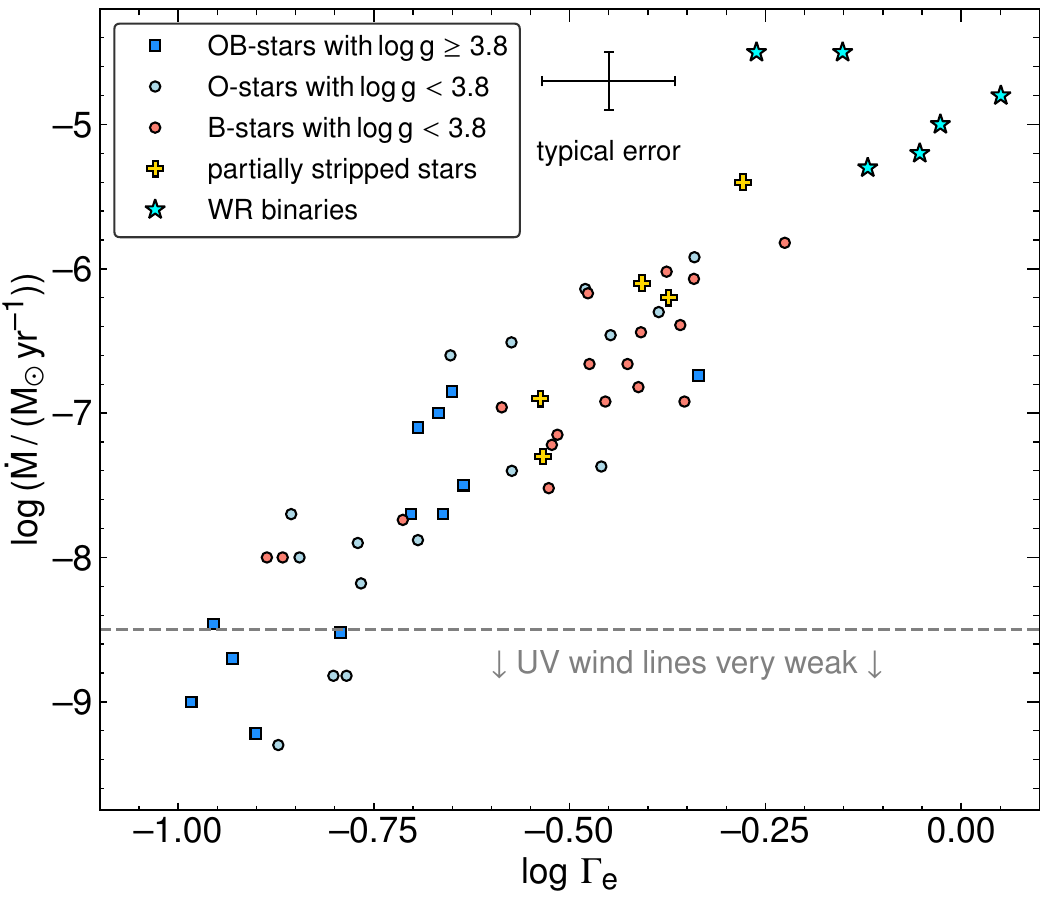}
     	    \caption{Mass-loss rates of SMC stars as a function of luminosity (left) and Eddington parameter (right). Blue squares are OB-type stars with surface gravity values $\log(g/(\mathrm{cm\,s^{-2}}))\geq3.8$ (i.e., main-sequence stars), light-blue circles are O-type stars with $\log(g/(\mathrm{cm\,s^{-2}}))<3.8$ (i.e., O-giants and O-supergiants), red circles are B-type stars with $\log(g/(\mathrm{cm\,s^{-2}}))<3.8$ (i.e., B-supergiants), yellow plus symbols are the post-interaction binaries recently discovered in the SMC, and cyan star symbols the Wolf-Rayet binaries. Note that for stars with mass-loss rates below ${\log(\dot{M}/(\msunpyr))\lesssim-8.5}$ the wind features in the UV get very weak, introducing larger uncertainties in their mass-loss rate estimates.}
     	    \label{fig:SMC_MDOT}
     	\end{figure*}

        Recently, a new class of post-interaction binaries has been identified in the SMC \citep{pau2:22,ric1:23,ram1:23,ram1:24}. Their detailed multi-wavelength and multi-epoch spectroscopic analyses revealed that the partially stripped stars in these systems have OB spectral type but exhibit mass-loss rates several orders of magnitude higher than those of ``normal'' OB-type stars with similar luminosities (see left panel of Fig.~\ref{fig:SMC_MDOT}). In fact, stellar winds of O-type partially stripped stars are in the same regime as winds of WR stars. 

        As previously noted, the luminosity-to-mass ratio, and as such the Eddington factor $\Gamma_{\!\text{e}}$, is of major importance in determining the strength of a stellar wind. When comparing the mass-loss rates of partially stripped stars to those of OB and WR stars as a function of $\Gamma_{\!\text{e}}$ (see right panel of Fig.~\ref{fig:SMC_MDOT}), rather than a function of $L$ (left panel of Fig.~\ref{fig:SMC_MDOT}), a notable trend emerges. This suggests that the strong winds of partially stripped stars are a consequence of their enhanced luminosity-to-mass ratios, and that a differentiation between the winds of OB-type, stripped, partially stripped non-WR and WR star is not needed.

        The recent discoveries of (fully) stripped intermediate-mass helium stars by \citet{dro1:23} show much weaker mass-loss rates than the partially stripped stars \citep{goe1:23}. Their stripped stars have lower Eddington parameters and the upper limits on their mass-loss rates align with the shown mass-loss relation. This implies that the low and high mass-loss rates of fully and partially stripped stars can be explained by their change in $\Gamma_{\!\text{e}}$ during their evolution.
        
    \subsection{Mass-loss rates of radiation-driven stellar winds}

\begin{figure*}[tbhp]
     	    \centering
     	    \includegraphics[width=\textwidth]{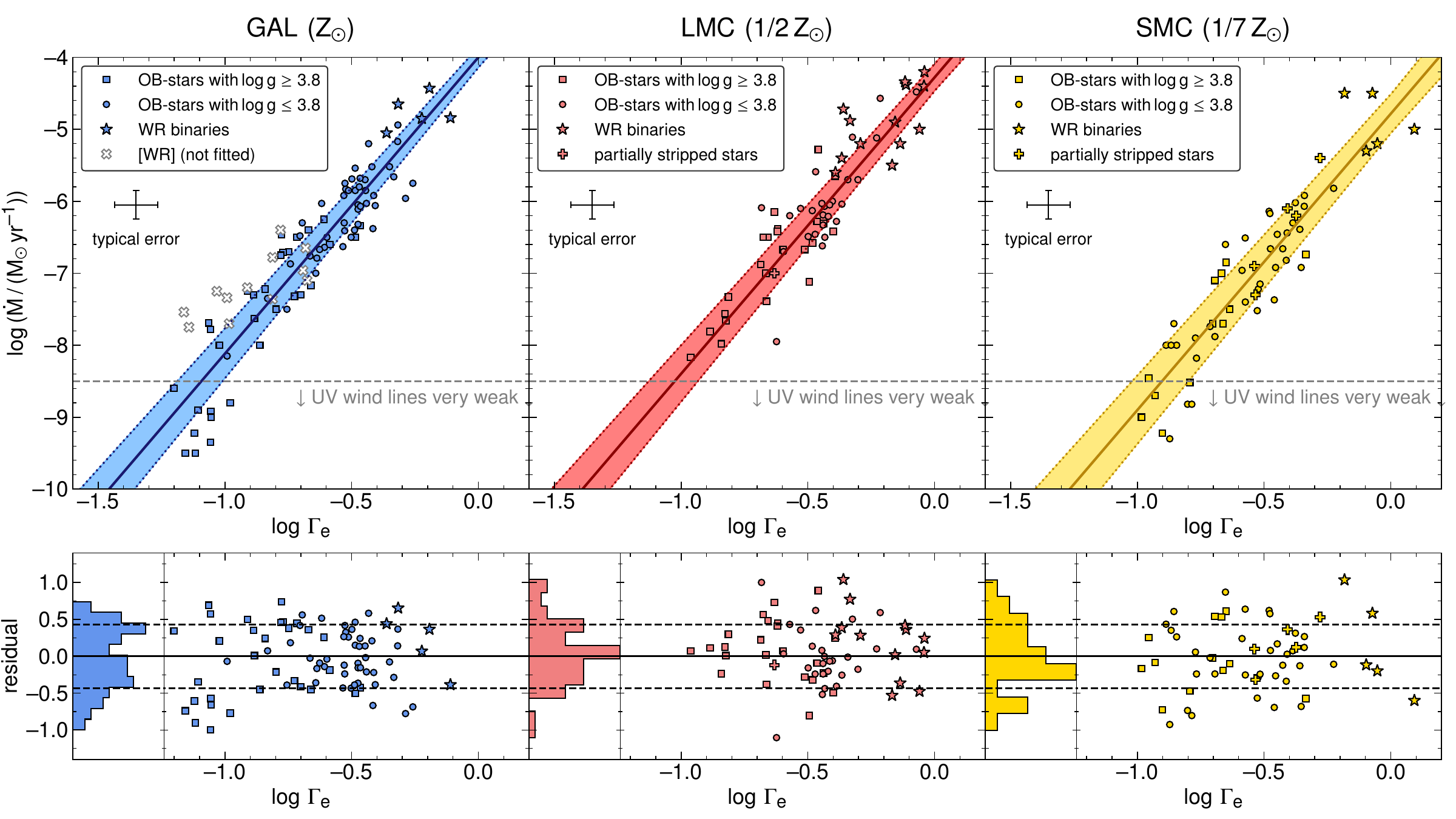}
            \caption{\textit{Upper:} Mass-loss rates of stars in the Galaxy (left), LMC (middle), and SMC (right) as a function of the Eddington parameter. The symbols indicate stars from different evolutionary stages as indicated in the legend. Solid lines and the shaded areas mark the best fit and the uncertainty, respectively. For the Galactic sample, the central stars of planetary nebulae ([WR]) stars are shown, but since their mass is not known and fixed to an average mass of $0.6\,\msun$ they are not included in the fit. \textit{Lower:} Residuals between the observed mass-loss rates and the best fit. A histogram illustrates the average spread around the mean value. Dashed lines indicate the $1\sigma$ root-mean-square dispersion from our fitting routine.}
     	    \label{fig:MDOT_GAMMAe}
     	\end{figure*}

        To further quantify the mass-loss rates of hot stars, we consider a sample of OB and WR stars of the Galaxy, LMC, and SMC (see Sect.\ref{sec:sample}). We find that the mass-loss rates of hot massive stars in all three galaxies also follow a power law when plotted against the Eddington parameter $\Gamma_{\!\text{e}}$ (see Fig.~\ref{fig:MDOT_GAMMAe}). While there is a considerable scatter in the sample, which we discuss below, the slope in $\log(\dot{M})\propto\log(\Gamma_{\!\text{e}})$ can be treated as universal for all three galaxies.

        The uncertainties on the measured stellar and wind parameters are not provided for some spectral analyses considered in this work. However, since the uncertainties in the measurements of the stellar and wind parameters are comparable among different spectroscopic analyses \citep{san1:24}, we decided to apply a uniform error estimate for the stars in our sample based on the typical error margins reported in the various literature studies included in our sample. For metallicity, we assume a precision of $10\,\%$, accounting for potential chemical gradients within the galaxies. By performing a $\chi^2$-fit to the complete dataset that considers the uncertainties in $\dot{M}$, $\Gamma_\text{\!e}$, and $Z_\text{Fe-group}$, we derived a universal relation for the mass-loss rates of hot stars as
        \begin{align}
            \log(\dot{M}/(\msunpyr))=   &+4.27(\pm0.13)\log(\Gamma_{\!\text{e}})\nonumber \\ 
                                        &+0.86(\pm0.09)\log(Z_\text{Fe-group}/\zsun)\nonumber \\
                                        &-3.92(\pm0.08).  \label{eq:mdotfit}
        \end{align}
        The lower panels of Fig.~\ref{fig:MDOT_GAMMAe} present the residuals between the empirically derived mass-loss rates and our fitted scaling relation (Eq.~(\ref{eq:mdotfit})), along with histograms showing $1\sigma$. The relations root-mean-square value of the residuals is ${\Delta\log(\dot{M}/(\msunpyr))=\pm0.43}$, implying that the mass-loss rates according to Eq.~(\ref{eq:mdotfit}) in most cases are as accurate within a factor of $2.7$. 
        
        The scatter around the scaling relation can be attributed to three factors. Firstly, the empiric mass-loss rates of our sample stars were compiled from the literature. These measurements have some inherent scatter due to different fitting methods and the use of different stellar atmosphere codes. Secondly, we do not account for factors that are expected to affect the mass-loss rate according to theoretical studies, such as $T_\text{eff}$ \citep[e.g.,][]{cas1:75,luc1:93}, rotational velocities \citep[e.g.,][]{fri1:86,mul1:14}, or CNO abundances \citep[e.g.,][]{abb1:82,san2:20}. Thirdly, using different empiric mass-loss diagnostics in the UV or optical can lead to an additional scatter in the derived mass-loss rates. Additionally, stellar winds are not stationary. Although we excluded stars that are known to have strong wind variability, cases, where only a single epoch spectrum is available, are retained in our sample, introducing an intrinsic scatter. Recently \citet{mas1:24} have demonstrated that any single mass-loss rate determination can differ from the mean by a factor of 2 or more due to wind variability. Thus the derived scaling relation in Eq.~(\ref{eq:mdotfit}) represents an average over these intrinsic uncertainties.
    
        Our sample includes stars that deviate significantly, with mass-loss rates up to a factor of 10 different from the scaling relation. These outliers include the Galactic stars with the lowest $\Gamma_{\!\text{e}}$ which are associated with the ``weak wind'' problem, where the majority of the wind material might be in higher ionization stages observable mainly in the X-ray regime rather than the optical or the UV \citep{luc1:12,hue1:12,lag1:21,law1:24}.
        
        Furthermore, some WR binaries exceed the $1\sigma$ threshold or have Eddington parameters larger than $\Gamma_{\!\text{e}}>1$. These systems have to be treated with care. For some WR binaries in the LMC, we adopted solutions where the orbital inclinations were derived by matching the spectroscopic and orbital masses of the secondary stars \citep[for more details see][]{she1:19}. This approach may introduce additional uncertainties in the derived masses and thus $\Gamma_{\!\text{e}}$, contributing to the observed scatter. The WR binaries in the SMC have orbital solutions with roughly constrained inclinations, leading to large uncertainties in the derived orbital masses which might be underestimated, and hence might have overestimated Eddington parameters. Additionally, the WR binary SMC\,AB\,5 (HD\,5980) is known for its complexity and variability, includes a luminous blue variable (LBV) companion that has recently undergone an eruptive phase \citep{koe1:04,koe1:14}. Caution is required when interpreting data from such systems. Removing these outliers from the fitting process does not significantly affect the quality of our linear fit, indicating that the overall power-law trend remains robust. 
        
        In the left panel of Fig.~\ref{fig:MDOT_GAMMAe}, where the mass-loss relation for the Galactic stars is displayed, we also include the positions of central stars of planetary nebulae ([WR]). These objects exhibit WR-like spectra but have much lower theoretical mass estimates, ranging between $0.4\msun\text{\,--\,}0.8\msun$ \citep{tyl1:03}. Despite their significantly lower wind efficiencies compared to classical WR stars -— and consequently lower $\Gamma_{\!\text{e}}$ -- they still align well with our derived mass-loss relation. This alignment supports the work from \citet{toa1:24} who suggest that the winds in these stars operate under similar physical principles as those in classical WR stars. This further supports the capability of our mass-loss relation in describing the UV-radiation-driven winds of hot stars.

    \section{Discussion}
    \label{sec:discuss}        

    \subsection{Metallicity dependence of radiation-driven winds}
    \label{sec:discuss_z}

        \begin{table}[tbp]
            \centering
            \caption{Metallicity dependence of the stellar winds of hot stars.}
            \small
            \begin{tabular}{cll}\hline \hline \rule{0cm}{2.8ex}%
                \rule{0cm}{2.2ex}   $m$     & Method            & Reference         \\
                \hline
                \multicolumn{3}{c}{\rule{0cm}{2.4ex} OB}\\
                \hline \rule{0cm}{2.4ex}%
                                    \rule{0cm}{2.4ex}0.94    & modified CAK      & \citet{abb1:82}   \\
                                    \rule{0cm}{2.4ex}0.5     & modified CAK      & \citet{kud1:87}   \\
                                    \rule{0cm}{2.4ex}0.8     & modified CAK      & \citet{pul1:00}   \\
                                    \rule{0cm}{2.4ex}0.85    & Monte Carlo calc.     & \citet{vin1:01}   \\
                                    \rule{0cm}{2.4ex}0.83    & empiric $\dot{M}$ (\Halpha\ only)     & \citet{mok1:07}   \\
                                    \rule{0cm}{2.4ex}\SIrange{0.5}{0.8}{}$^\dagger$    & empiric $\dot{M}$ (UV+opt.)    & \citet{mar1:22}   \\
                                    \rule{0cm}{2.4ex}\SIrange{0.66}{1.64}{}$^{\dagger\dagger}$     & hydrodynamic stellar   & \citet{bjo1:23}   \\
                                    &   atmosphere models&\\
                \hline
                \multicolumn{3}{c}{\rule{0cm}{2.4ex} WN}\\
                \hline \rule{0cm}{2.4ex}%
                                    \rule{0cm}{2.4ex}0.86     & Monte Carlo calc.      & \citet{vin1:05}   \\
                                    \rule{0cm}{2.4ex}$1.2\pm0.1$     & empiric $\dot{M}$      & \citet{hai1:17}   \\
                                    \rule{0cm}{2.4ex}$0.83\pm0.09$   & empiric $\dot{M}$      & \citet{she1:19,she1:20}   \\
                                    \rule{0cm}{2.4ex}$1.3\pm0.2$     & empiric $\dot{M}$      & \citet{tra1:16}   \\
                \hline
                \multicolumn{3}{c}{\rule{0cm}{2.4ex} WC/WO}\\
                \hline \rule{0cm}{2.4ex}%
                                    \rule{0cm}{2.4ex}0.66     & Monte Carlo calc.      & \citet{vin1:05}   \\
                                    \rule{0cm}{2.4ex}$0.25\pm0.08$   & empiric $\dot{M}$      & \citet{tra1:16}   \\
                \hline
                \multicolumn{3}{c}{\rule{0cm}{2.4ex} OB+WR}\\
                \hline \rule{0cm}{2.4ex}%
                                    \rule{0cm}{2.4ex}0.86    & empiric $\dot{M}$  & this work  \\
                \hline
            \end{tabular}
            \rule{0cm}{2.8ex}%
            \begin{minipage}{0.95\linewidth}
                \ignorespaces                  
                \rule{0cm}{2.8ex}$^\dagger$ The authors report that the metallicity dependence is a function of luminosity. The quoted exponents refer to stars with luminosities above $\log(L/\lsun)\gtrsim5.4$ (i.e. valid for most of our stars). \\$^{\dagger\dagger}$The authors predict that the metallicity dependence is a function of temperature. We calculated the metallicity exponents for the provided validity regime of their relation.
                 
            \end{minipage}
            \label{tab:exp_OB}
        \end{table}

        Understanding the metallicity dependence of UV-radiation-driven winds is of major importance for understanding stars formed in the young Universe. Over the years several studies focused on deriving the metallicity dependence of OB-type star winds, yielding values in between $m=\SIrange{0.5}{1.64}{}$, centering around a value of $m\approx0.85$ (see Table~\ref{tab:exp_OB}). The metallicity exponent of $m=0.86\pm0.09$ derived in this work aligns well with these previous results. However, a critical distinction lies in the assumed metallicity of the SMC. For instance, in this work, we assume that the winds of massive stars are driven by the iron group elements, which yields ${Z_\mathrm{SMC,\,\text{Fe-group}}=1/7\,\zsun}$. In the work of \citet{mok1:07} and \citet{mar1:22} only the iron abundance is considered yielding ${Z_\mathrm{SMC,\,\text{Fe}}=1/5\,Z_{\odot,\,\text{Fe}}}$. If we only consider a metallicity scaling with iron, our derived metallicity exponent increases to $m=1.03\pm0.11$. This highlights the sensitivity of metallicity-dependent wind scaling laws to the adopted metallicity of a host galaxy. Hence, when incorporating our mass-loss relation into stellar evolution codes it is recommended to adjust the metallicity scaling to the assumed iron-group abundance relative to the solar value.
        
        For the optically thick winds of WR stars metallicity exponents in the range from $m=\SIrange{0.25}{1.3}{}$, centering -- as for OB-type stars -- around $m\approx0.85$ (see Table~\ref{tab:exp_OB}), are reported. The metallicity trend of WN-type stars agrees well with our derived metallicity dependence. However, previous studies focusing on WC/WO type stars report noticeably lower metallicity exponents which are out of line with our finding, but approximately agree with the asymptotic behavior of WR stars in hydrodynamically consistent calculations \citep{san1:20,san2:24}. Due to the constraints of our sample and selection criteria, only very few WC and WO type stars are included in this work, such that we cannot investigate whether similar trends for individual WR subtypes exist in our data. 
        
        Recently, \citet{san1:20} calculated hydrodynamically consistent stellar atmosphere models for WR stars which suggest that the metallicity dependence of UV-radiation-driven winds might not follow a simple power-law relationship. Instead, their models predict that WR star mass-loss rates sharply decline below the metallicity of the SMC. For OB-star winds, \citet{kud1:02} also found a breakdown of the power-law behavior for very low $Z$. Given that our sample is to the nearby galaxies and the SMC provides the lowest metallicity anchor, we advise caution when extrapolating mass-loss relations to extremely low metallicities.

    \subsection{The sharp increase in the mass-loss rate for stars near the Eddington limit}

        The classical CAK theory predicts that stars approaching the Eddington limit experience a significant increase in their mass-loss rates, resulting in a dramatic boost in $\dot{M}$ (see Eq.~(\ref{eq:CAK})). This effect is also predicted in the semi-analytic prescription of \citet{bes1:20}, who suggest a transition occurring at $\log(\Gamma_{\!\text{e}}) \gtrsim -0.3$ for stars in the LMC. 

        From Monte Carlo simulations of stellar winds, \citet{vin1:11} predict that the transition from optically thin to optically thick winds introduces a shift in the driving mechanism of the wind, resulting in a different $\dot{M}$-$\Gamma_{\!\text{e}}$ relation. Specifically, their models suggest a ``kink'' in the $\dot{M}$-$\Gamma_{\!\text{e}}$ relation. Supporting this theory, the empiric study by \citet{bes1:14} reported evidence for such a kink beginning at $\log(\Gamma_{\!\text{e}}) \gtrsim -0.6$ at LMC metallicity. However, their mass estimates for WR stars are derived using mass-luminosity relations based on chemically homogeneous stellar models, which may deviate from the orbital masses. 
        
        Additionally, \citet{vin1:12} derived a model-independent transitional mass-loss rate at which stellar winds are expected to shift from being optically thin to optically thick, and, hence, a kink should occur. For a Galactic O4-6If+ star with $\log(L/\lsun)=6.05$ and $M=60,\msun$ (corresponding to $\log(\Gamma_{\!\text{e}})\approx-0.3$), they proposed a transitional mass-loss rate of $\log(\dot{M}_\mathrm{trans}/(\msunpyr))=-4.95$. When comparing this prediction to our Galactic mass-loss relation (shown in the left panel of Fig.~\ref{fig:MDOT_GAMMAe}), one can see that stars with $\log(\Gamma_{\!\text{e}})\approx-0.3$ are indeed primarily evolved objects such as supergiants and WR stars, which have optically thick winds, thereby supporting their theoretical framework. However, our derived mass-loss rate at this Eddington parameter is approximately a factor of two lower than the transitional mass-loss rate proposed by \citet{vin1:12} but still lies within the scatter of our sample. 

        When inspecting Fig.~\ref{fig:MDOT_GAMMAe}, one cannot see any of the predicted sharp increases in mass-loss rates for stars close to the Eddington limit in any of the galaxies studied in this work. However, given the small number of stars with such high Eddington parameters in our literature sample, such a kink or upturn might be hidden in the large uncertainties of the Eddington parameters of the WR stars originating from the uncertainties of the orbital masses when the inclination is not well known. Only larger samples of stars near the Eddington limit with securely known $\Gamma_{\!\text{e}}$ could firmly confirm or rule out the presence of the theoretically predicted non-monotonic increase in mass-loss rates in stars close to the Eddington limit. Following this discussion, we advise caution when using our mass-loss relation for stars near the Eddington limit, as the behavior of winds in this regime remains uncertain and might deviate from the general trends observed in our dataset.
    
    \subsection{Comparison to other commonly used mass-loss recipes}
    \label{sec:discuss_mdot}

        Several studies have demonstrated that the choice of mass-loss prescriptions has profound implications for the evolution of massive stars \citep[e.g.,][]{ren1:17,gil1:19,sab2:22,jos1:24}. Many widely used mass-loss recipes depend on various stellar parameters rather than explicitly on the Eddington parameter, as in our newly derived prescription.

        To assess the differences and implications of our $\dot{M}-\Gamma_{\!\text{e}}$ relation compared to other mass-loss recipes, we computed stellar evolution models. These models cover the evolution of stars from the main sequence to the supergiant phase, as well as for pure helium (He) stars, spanning the full $\Gamma_{\!\text{e}}$ range of our sample. The models were generated using the MESA code, and details regarding the input physics and parameters are provided in Appendix~\ref{app:mesa}.
        
    \subsubsection{Mass-loss rates of OB-type stars}

     	\begin{figure*}[tbhp]
     	    \centering
     	    \includegraphics[trim={0cm 0cm 0cm 13.5cm},clip,width=\textwidth]{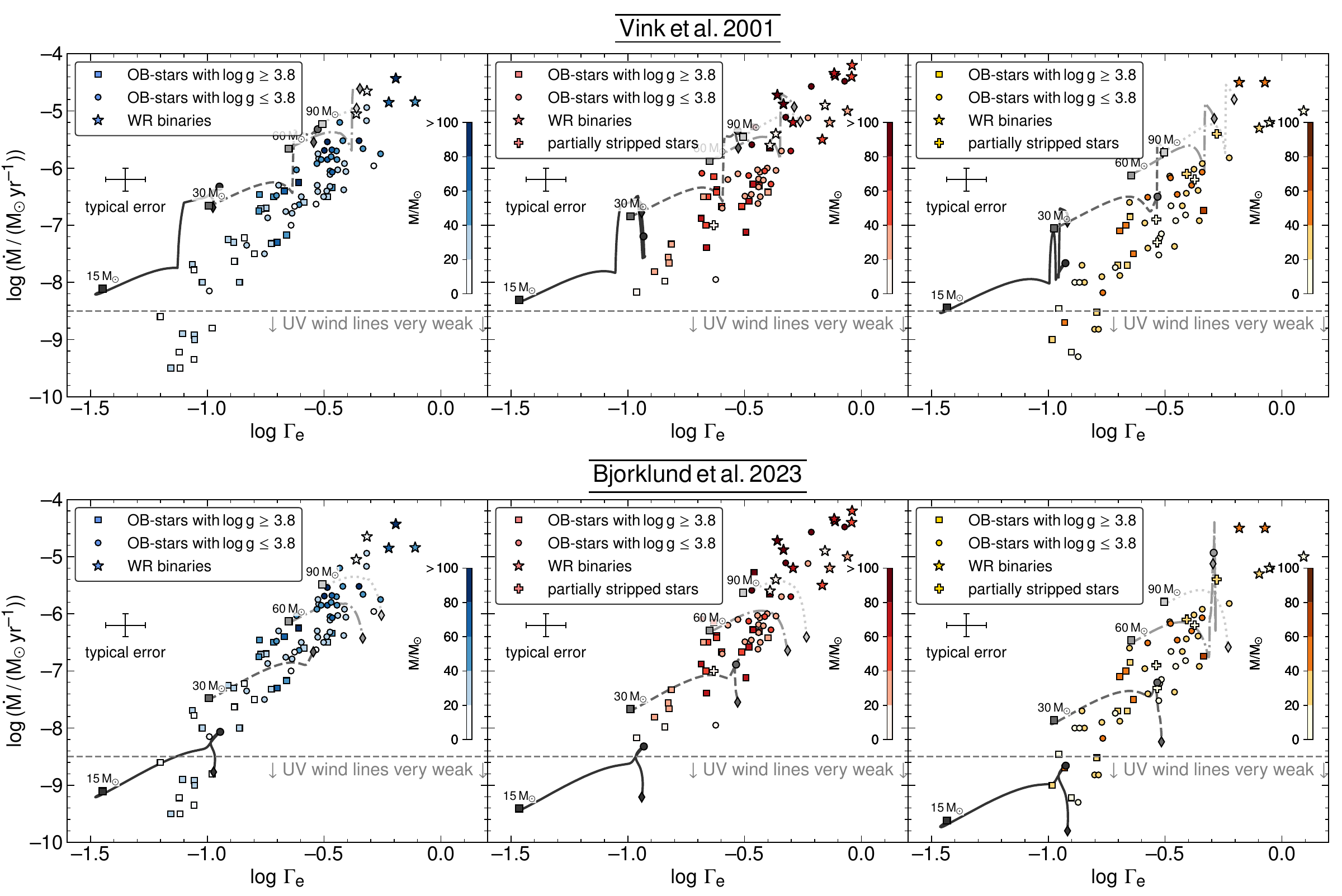}
     	    \caption{Mass-loss rates of stars in the Galaxy (left), LMC (middle), and SMC (right) as a function of the Eddington parameter. The discrete symbols indicate stars from different spectral classes. The symbols are color-coded according to a star's current mass. To enable a comparison to different mass-loss recipes (different rows) that do not have a direct dependence on $\Gamma_{\!\text{e}}$, we calculated stellar evolution models that predict how $\Gamma_{\!\text{e}}$ changes during a star's evolution. Lines indicate how $\dot{M}$ and $\Gamma_{\!\text{e}}$ change in the stellar evolution models with initial masses of $M_\mathrm{ini}=15\,\msun$ (solid line), $30\,\msun$ (dashed line), $60\,\msun$ (dash-dotted line), and $90\,\msun$ (dotted line) using different mass-loss recipes. The models shown here are calculated until either the surface hydrogen abundance drops below $X_\mathrm{H}<0.7$ (i.e., when the models would interpolate between MS and WR wind) or until a temperature below $T<\SI{12}{kK}$ (i.e., when the models would switch to an RSG wind) is reached. Hence, these models should not be compared to the classical WR stars. The different symbols on the lines mark the zero-age main sequence (squares), the terminal-age main sequence (circles), and the aforementioned stopping criterion (diamond).}
     	    \label{fig:mdot_ms}
     	\end{figure*}

        The most commonly used mass-loss recipe employed in stellar evolution calculations during the OB star phase is the one by \citet{vin1:01}. According to this recipe, mass-loss rates do not only depend on mass and luminosity but also temperature, as the latter determines the dominant ion responsible for driving the wind. In specific situations, such an ionization change is predicted to considerably boost the mass-loss rate \citep[see][]{vin1:99}, yielding a ``jump'' in the resulting curve when reaching specific temperature regimes. In the upper panels of Fig.~\ref{fig:mdot_ms}, we depict the change of $\dot{M}$ as a function of $\Gamma_{\!\text{e}}$ for stellar evolution models of main-sequence stars to the supergiant stage calculated with the \citet{vin1:01} mass-loss recipe to the empiric $\dot{M}-\Gamma_{\!\text{e}}$ plane. For massive Galactic stars, the predicted mass-loss rates show some overlap with the observations but still seem to be systematically higher. This discrepancy is even more evident at lower stellar masses and for stars in low-metallicity environments. Here, the \citet{vin1:01} mass-loss rates are consistently up to an order of magnitude higher than the observed values for stars with comparable properties. Additionally, the predicted temperature-dependent jumps in mass-loss rates, are not reflected in the empiric $\dot{M}-\Gamma_{\!\text{e}}$ plane, indicating that the \citet{vin1:01} prescription may overestimate these effects or that the physical assumptions underlying these jumps do not hold uniformly across the stellar population.

     	\begin{figure*}[tbhp]
     	    \centering
     	    \includegraphics[trim={0cm 0cm 0cm 13cm},clip,width=\textwidth]{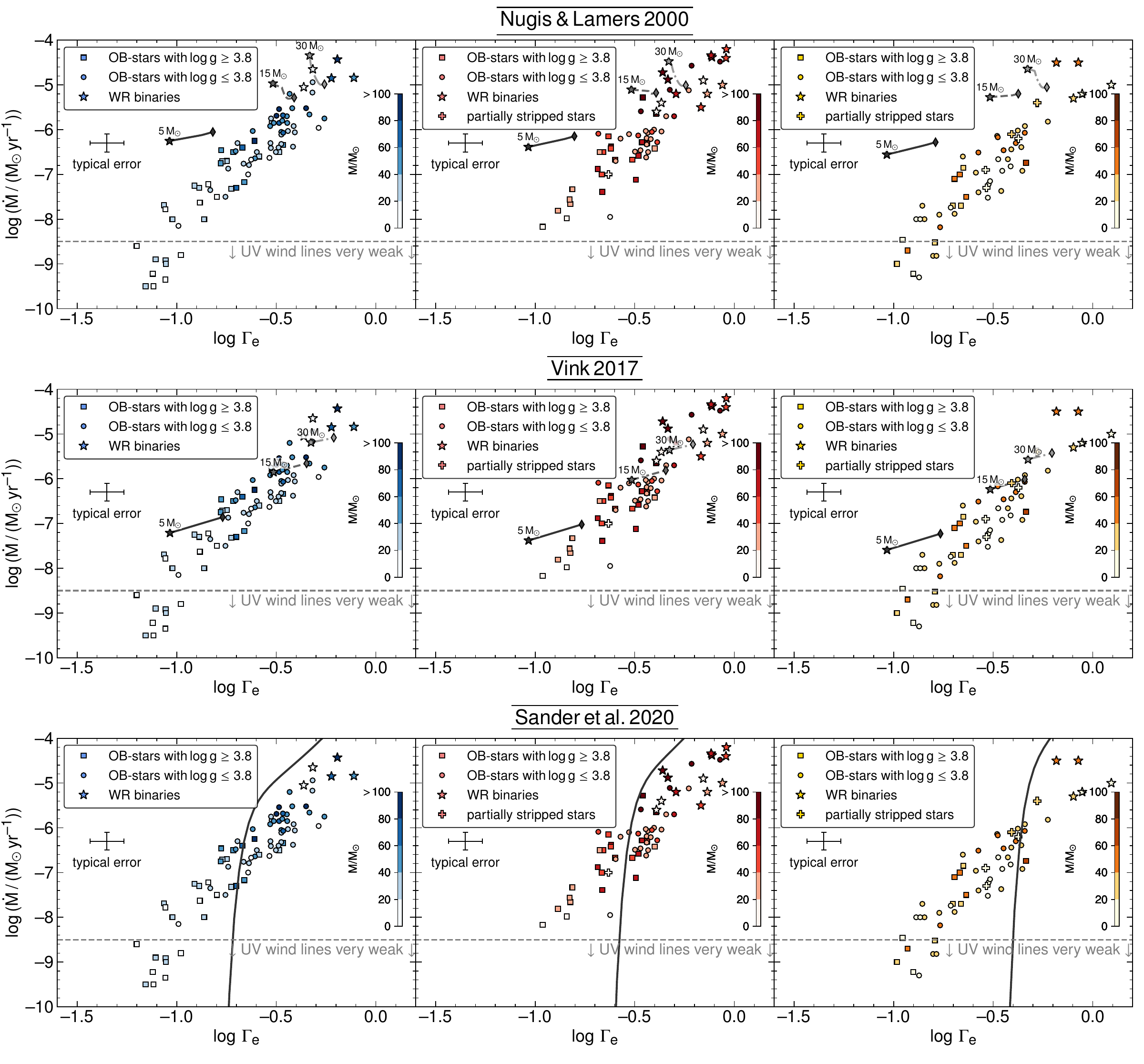}
     	    \caption{Same as Fig.~\ref{fig:mdot_ms}, but now for He-star models with initial mass $M_\mathrm{ini}=5\,\msun$ (solid line), $15\,\msun$ (dashed line), and $30\,\msun$ (dash-dotted line). The different symbols on the track mark the helium zero-age main sequence (stars), and core-helium depletion (diamond). For the lowest panels, the mass-loss recipe of \cite{san1:20}, which has a direct $\dot{M}\propto\Gamma_{\!\text{e}}$ scaling relation is shown as a solid line.}
     	    \label{fig:mdot_wr}
     	\end{figure*}
        
        Recently, \citet{bjo1:23} calculated dynamically consistent stellar atmosphere models providing new theoretical mass-loss rates for massive OB-type stars. Their models do predict a temperature dependence, but without jumps in the mass-loss rates. In the lower panels of Fig.~\ref{fig:mdot_ms}, we compare stellar evolution models incorporating their mass-loss prescription to the empiric $\dot{M}-\Gamma_{\!\text{e}}$ plane. The models generally align well with the empiric mass-loss rates across all metallicities. However, the models exhibit a notable trend: when a star expands and its surface temperature decreases during the supergiant phase, the predicted mass-loss rates tend to decline. One exception is seen in the $60\,\msun$ SMC model at its terminal-age main sequence (TAMS), where it experiences two sign flips that briefly enhance its mass-loss rate. If this trend of decreasing mass-loss rate for later evolutionary stages were strictly followed, one would expect a clear distinction along the $\dot{M}-\Gamma_{\!\text{e}}$ plane, namely that main-sequence stars (with $\log(g/(\si{cm\,s^{-2}}))\geq3.8$)) should cluster in the upper-left region, while supergiants ($\log(g/(\si{cm\,s^{-2}}))<3.8$) should occupy the lower-right. However, this distinction is not apparent in any of the three galaxies studied. It is unclear whether this discrepancy arises from intrinsic scatter and observational uncertainties or from unresolved physical processes in the theoretical models. A deeper investigation, potentially incorporating more precise observational constraints and refined modeling, is needed to resolve this question.
        
    \subsubsection{Mass-loss rates of He- and WR-stars}
    \label{sec:WRmdot}

        To compare different mass-loss recipes for He- and WR-star winds with observations and our newly derived relation, we calculated models of pure He stars. In Fig.~\ref{fig:mdot_wr}, we present how the He-star models evolve in the $\dot{M}-\Gamma_{\!\text{e}}$ plane in dependence on different mass-loss recipes. Note that most of the He- and WR stars in our sample are not completely free of H at the surface. Adding H to the surface increases $\Gamma_{\!\text{e}}$, so from our recipe, these H-rich WR stars should have mass-loss rates higher than those predicted by the pure He-star models.

        To model the He- and WR-star phases in stellar evolution calculations, often the mass-loss prescription from \citet{nug1:00} is applied. The upper panels in Fig.~\ref{fig:mdot_wr} illustrate how $\dot{M}$ changes as a function of $\Gamma_e$ during the evolution of a He-star model using this recipe. By comparison to the empiric $\dot{M}-\Gamma_{\!\text{e}}$ plane, it is evident that the mass-loss rates of \citet{nug1:00} are significantly overestimated, being about a factor of three higher than observed for Galactic stars. This mismatch is surprising, given that the rates from \citet{nug1:00} were calibrated using Galactic stars and are based on radio data with clumping correction. They are also roughly in line with the results from \citet{ham1:06,san1:19}. Assuming this difference is not due to any diagnostic issues, the apparent offset could be due to the difference between the WR sample selected for this work and the underlying sample of \citet{nug1:00}. Except for WR137, all of the Galactic WR stars selected for this work have rather small mass-loss rates and three out of the five stars are high-mass WNh-type stars. The discrepancy between our evolution models using \citet{nug1:00} and the curated observational sample becomes even more pronounced at lower metallicities due to the relatively shallow metallicity dependence in the \citet{nug1:00} prescription, $\dot{M}\propto Z^{0.5}$. 
        
        However, one has to be careful when comparing their mass-loss rates across metallicity to our results, as their metallicity includes all metals, while our metallicity only refers to the iron content of a star, which is not expected to change during the evolution. When extrapolated to intermediate-mass He stars ($5\,\msun$), the mass-loss rates predicted by \citet{nug1:00} are at least an order of magnitude too high. Such strong mass-loss rates can have substantial consequences for stellar evolution models, as they heavily influence the evolutionary pathways and final fates of stars, particularly in determining whether a star ends its life as a neutron star or black hole. 

        Another mass-loss relation often considered for helium stars, especially for those classified as ``non-WR'' (i.e., He-stars without the characteristic spectroscopic signatures of WR stars), is the prescription from \citet{vin1:17}. Assuming a fixed temperature of $T = \SI{50}{kK}$, this recipe was purposely developed for non-WR He-stars. 
        The corresponding He-star models in the $\dot{M}-\Gamma_{\!\text{e}}$ plane are illustrated in the second row of panels in Fig.~\ref{fig:mdot_wr}. Despite its simplicity -- being a function solely of luminosity and metallicity -- the \citet{vin1:17} mass-loss rates for He-stars align remarkably well with the weaker part of the observed mass-loss rates of WR stars across the Galaxy, LMC, and SMC. Moreover, it agrees roughly with the mass-loss rates of partially stripped stars included in the analysis. As discussed above, our He star models do not reflect the higher $\Gamma_{\!\text{e}}$ of the partially stripped stars and thus there is a mismatch between the $5\,M_\odot$ model and partially stripped stars in a similar mass regime. Despite the general agreement of the \citet{vin1:17} description, one has to be careful in its applications in evolution models given that it contains only a direct $L$ and no $\Gamma_{e}$-dependence. Applying this mass-loss relation outside its intended parameter space could lead to significantly different mass-loss rates and thus deviations in evolutionary predictions. 

        \citet{san2:20} developed hydrodynamically consistent stellar atmosphere models for He-stars, covering a broad range of initial masses and metallicities. Their work suggests that the mass-loss rates of He- and WR-stars can be described with a $\Gamma_{\!\text{e}}$-dependence, which aligns conceptually with our findings. The lower panels of Fig.~\ref{fig:mdot_wr} show a comparison between the updated \citet{san1:20} theoretical mass-loss prescription and our sample of WR and He-stars. 
        One can see that the mass-loss rates predicted by \citet{san1:20} are about an order of magnitude higher than the average rates in our sample, but seem to follow a similar trend. The mass-loss recipe of \citet{san1:20} has a $\log(\dot{M})\propto \log(-\log(1-\Gamma_\text{\!e}))$ dependence, making a direct comparison to our empirical mass-loss relation difficult. When assuming that $\log(1+x) \approx x$, their mass-loss recipe for stars with $\Gamma_\text{\!e}\ll1$ can be approximated by $\log(\dot{M})\propto \log(\Gamma_\text{\!e})$. Following their theoretical calculations, they derive a slope of $2.93\pm0.02$ \citep[][their equation~29]{san1:20}, which is somewhat shallower than our empirically derived slope of $4.18\pm0.18$ (Eq.\,\eqref{eq:mdotfit}).
                
    \subsection{Implications for stellar evolution}
    \label{sec:mesa}

     	\begin{figure*}[tbhp]
     	    \centering
     	    \includegraphics[width=\textwidth]{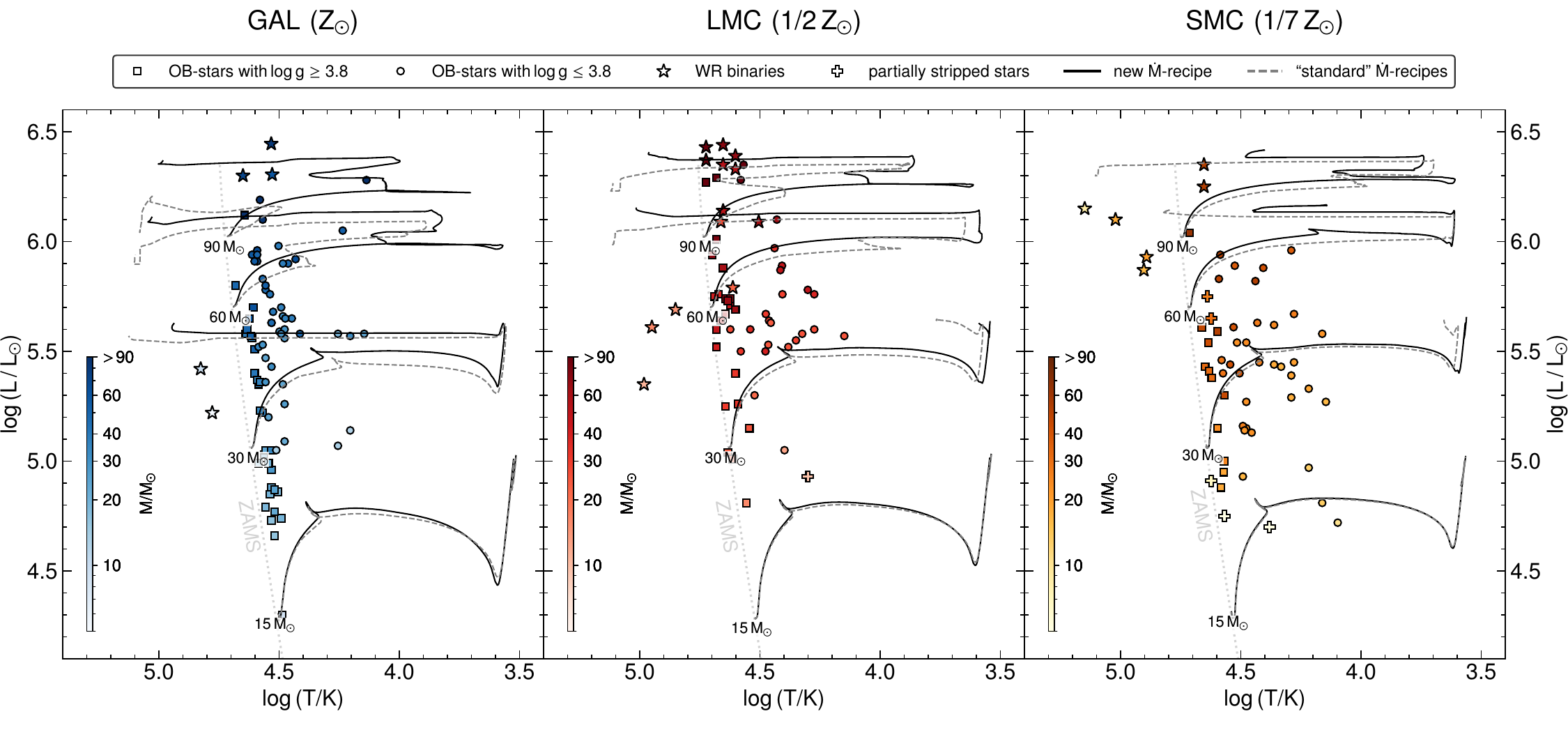}
     	    \caption{HRD containing the positions of the stars in our sample for the Galaxy (left), LMC (middle), and SMC (right) over-plotted by stellar evolution tracks using our mass-loss recipe (solid lines) and the standard wind mass-loss rates (dashed lines) and an initial rotation velocity of $\varv_\mathrm{rot,\,ini}=100\,\mathrm{km\,s^{-1}}$. The individual stars are color-coded by their current mass.}
     	    \label{fig:hrd_new_mdot}
     	\end{figure*}
        
        To explore the impact of our newly derived mass-loss recipe on stellar and binary evolution, we computed stellar evolution tracks for stars at solar (GAL), LMC, and SMC metallicities. These tracks employed two different mass-loss paradigms: a ``standard'' set of recipes commonly used in stellar evolution calculations, and our newly derived mass-loss rates for stars with radiative envelopes. This investigation aims to highlight the consequences of adopting a $\Gamma_{\!\text{e}}$-scaling on stellar evolution, particularly in low-metallicity environments, rather than creating a perfectly realistic stellar population.
        
        As the set of standard mass-loss recipes, we used the \citet{vin1:01} prescription for OB-type stars with surface temperatures $T>\SI{12}{kK}$ and surface H-abundances $X_\mathrm{H}>0.7$. For the He- or WR-stage, the \citet{nug1:00} mass-loss rates are employed as soon as the surface H-abundances drop below $X_\mathrm{H}<0.4$. We linearly interpolate between the two mass-loss recipes for transition phases between the OB and WR stages. To model red supergiant mass-loss (RSG), we employ the prescription of \citet{nie1:90}  for stars cooler than $T<\SI{10}{kK}$. For models with temperatures between $\SI{12}{kK}>T>\SI{10}{kK}$ we linearly interpolate between the hot star mass-loss rates to the RSG ones to avoid sudden jumps.
    
        For the ``new'' stellar evolution models, we use our new prescription for stars hotter than $T>\SI{12}{kK}$ and the RSG mass-loss rates from \citet{nie1:90} for models with temperatures below $T<\SI{10}{kK}$. We again interpolate between the two mass-loss recipes for stars with temperatures in between $\SI{12}{kK}>T>\SI{10}{kK}$.
        
        The other input physics are kept fixed between the two sets of stellar evolution models and try to mimic the setup as described by \citet{bro1:11}. Full details on the assumed input physics are provided in Appendix~\ref{app:mesa}.
      
    \subsubsection{Single star evolution}

        In Fig.~\ref{fig:hrd_new_mdot}, we present HRDs showing our literature sample alongside stellar evolution tracks for solar (GAL), LMC, and SMC metallicities, calculated using the two different setups for the mass-loss rates. The differences between the tracks and their final evolution for each galaxy are detailed below.

        In the case of the Galactic models, all stars with initial masses above $M_\mathrm{ini}\gtrsim30\,\msun$ evolve into WR stars, irrespective of the adopted recipes. However, notable differences emerge when comparing the evolutionary path that led to their final evolutionary stage. For instance, in the case of the standard mass-loss recipes, the $90\,\msun$ model undergoes significant mass loss during the main sequence, which rapidly removes the hydrogen-rich envelope. This process reduces the surface hydrogen abundance below $X_\mathrm{H}<0.7$, leading to a direct transition into a WR star without an intermediate expansion toward the red supergiant (RSG) region. Consequently, this mass-loss recipe struggles to explain the observed luminous supergiants, as these stars skip the evolutionary phase that places them in this region of the HRD. In contrast, the $90\,\msun$ model using our new mass-loss recipe retains its H-rich envelope for a longer duration, allowing the star to expand temporarily into the RSG region, penetrating for a short time the empirical Humphreys-Davidson (HD) limit, before shedding its H-rich envelope and becoming a WR star. This evolution aligns better with the presence of luminous supergiants in the observed sample. However, the new recipe also predicts WR stars with luminosities above $\log(L/L_\odot)>6.0$. Another major difference between the two sets of stellar evolution models is the different WR subtypes they produce. For the models using the standard mass-loss recipes, the $60\,\msun$ and the $90\,\msun$ model both will end their lives as WC-type stars, while for the models using our new mass-loss recipe, they will end their life only as H-free WN type stars.

        For the LMC models, the differences between the two mass-loss recipes are less pronounced than for the Galactic case but remain notable. For the $15\,\msun$ and $30\,\msun$ models, the differences in their evolution are only marginal. For the more massive models with initial masses of $60\,\msun$ and $90\,\msun$ differences still persist. For instance, for the models using the standard mass-loss recipe, the winds are very efficient and lead to a switch to the WR winds during the main sequence. Such a jump cannot be seen in the models using our new mass-loss recipe, which expands further during core-H burning. Nevertheless, in both models, with the new and the standard mass-loss rates, all stars expand to an RSG phase and all models penetrate the HD limit for a noticeable time. During this time the stellar models efficiently lose their H-rich envelope, allowing them to become WR stars. Similarly to the Galactic case, the models with our new mass-loss recipe end their lives as WN-type stars, while the models with the standard mass-loss recipe will finish their evolution as WC-type stars. In both cases, the predicted RSGs and WRs exhibit luminosities exceeding those observed. 

        At SMC metallicity, the weak mass-loss rates have minimal influence on the evolution of stars with initial masses of $15\,\msun$ and $30\,\msun$. Even at higher initial masses, the differences during the main sequence are small. The models using the new mass-loss recipe lose less mass during their evolution and thus inflate less and reach their terminal age main sequence at higher temperatures. Similar to the LMC models, most of the H-rich envelope is lost during an RSG phase. Given that at this metallicity the mass-loss rates of our recipe are about one order of magnitude lower than the ones from \citet{nug1:00} (see also Sect.~\ref{sec:WRmdot}), the single star models using our mass-loss rates never fully strip and do not become He- or WR stars.

        This comparison highlights how adopting our new mass-loss recipe influences a star’s ultimate fate. Different mass-loss rates result in different core-to-envelope ratios, affecting the stars' internal structures and subsequent evolution. This is particularly important in the high-mass regimes where an inefficient stripping of an envelope can lead to a different composition at core collapse, changing the final supernova from type Ib/c to IIb or even II. This greatly impacts the mechanical feedback and chemical enrichment associated with the supernova, hence its impact on a galaxy's evolution. Furthermore, having in general lower mass-loss rates could potentially reduce the longstanding mass discrepancy problem, where spectroscopic masses inferred from observations often do not match evolutionary masses derived from stellar models. Nevertheless,
        we emphasize that 
        besides $\dot{M}$ 
        it is crucial to understand the mixing processes within the stars as well as to incorporate updated mass-loss rates for RSGs \citep[see, e.g.,][]{zap1:24}, and eruptive mass loss \citep[e.g., as discussed by][]{lan1:94,mae1:97,eks1:12,vin1:23,che1:24}.

    \subsubsection{Binary evolution}

        \begin{figure*}[tbhp]
     	    \centering
     	    \includegraphics[trim={0cm 1.26cm 11.5cm 0cm},clip,width=0.8\textwidth]{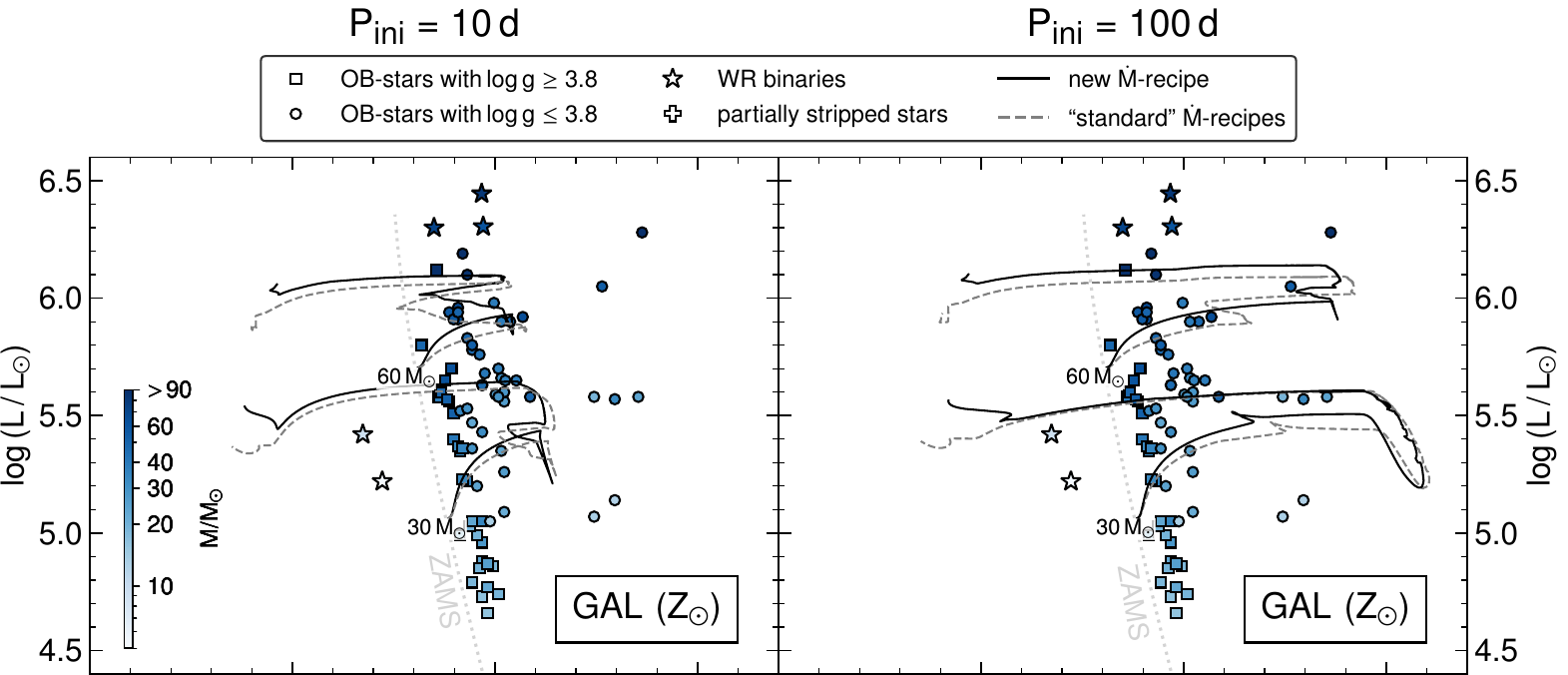}
     	    \includegraphics[trim={0cm 1.26cm 11.5cm 2.67cm},clip,width=0.8\textwidth]{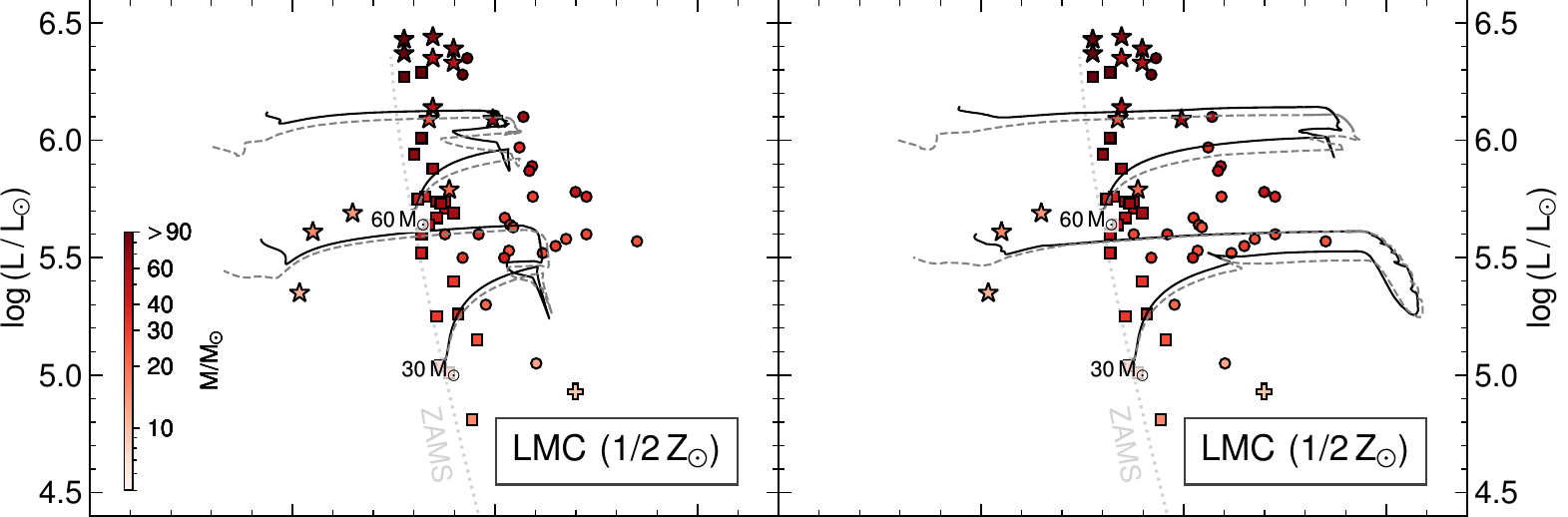}
     	    \includegraphics[trim={0cm 0cm 11.5cm 2.67cm},clip,width=0.8\textwidth]{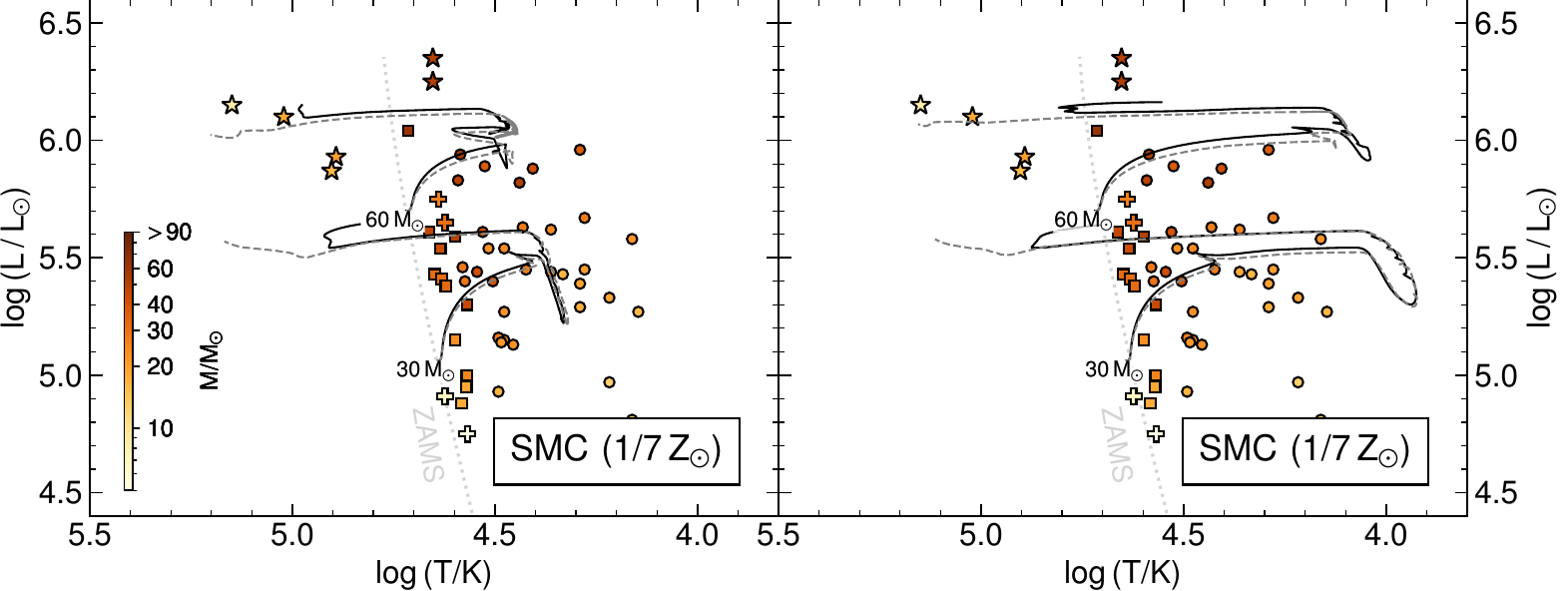}
     	    \caption{HRDs containing the positions of the stars in our sample for the Galaxy (top), LMC (middle), and SMC (bottom) over-plotted by binary evolution tracks of the primaries with initial orbital periods of $P_\mathrm{ini}=\SI{10}{d}$ (left) and $P_\mathrm{ini}=\SI{100}{d}$ and a fixed mass-ratio of $q_\mathrm{ini}=0.8$. Tracks calculated using our mass-loss recipe and the standard wind mass-loss rates are shown as solid and dashed lines, respectively. The individual stars are color-coded by their current mass.}
     	    \label{fig:hrd_new_mdot_binary}
     	\end{figure*}

        Additionally, we performed test calculations of
        binary evolution models with primary stars of initial masses $M_1=30\,\msun$ and $60\,\msun$, a fixed initial mass ratio of $q_\mathrm{ini}=0.8$, and initial orbital periods of $P_\mathrm{ini}=\SI{10}{d}$ and $P_\mathrm{ini}=\SI{100}{d}$ (i.e., Case A and Case B systems). Figure~\ref{fig:hrd_new_mdot_binary} displays the evolutionary tracks of the primary stars using the new and standard mass-loss rates.
        
        In our models, we generally see that primaries in shorter-period systems lose more of their hydrogen-rich envelopes during mass-transfer events. This leads to lower final masses, with approximately $1\,\msun\text{\,--\,}2\,\msun$ difference in the final stellar masses for the models with $P_\mathrm{ini}=\SI{10}{d}$ and $P_\mathrm{ini}=\SI{100}{d}$. 

        In the set of the Galactic models, the primaries using the standard mass-loss recipe lose significantly more mass on the main sequence (lower luminosity and terminal age main sequence at lower temperature) compared to the models calculated using our mass-loss rates, meaning that there is already a noticeable difference in mass when these stars enter the mass-transfer phase. While the tracks of models with different mass-loss prescriptions can appear similar after the mass-transfer phase, the physical properties of the primaries are distinct. Models using the new mass-loss rates result in primaries that are a few solar masses more massive after mass transfer compared to those calculated with the standard mass-loss rates, emphasizing the long-term impact of early evolutionary stages on the final fate of the star.

        After mass transfer, the $30\,\msun$ primaries calculated with the standard mass-loss recipe efficiently lose most of their H-rich envelopes and rapidly transition to the He-zero-age main sequence (He-ZAMS). In contrast, primaries calculated with the new mass-loss recipe lose less mass and temporarily remain between the He-ZAMS and H-ZAMS before becoming H-free WN-type stars, which might help to explain the observed spread in temperature of WR stars. 
        
        For the $60\,\msun$ primaries, the divergence between the tracks is more pronounced. Models with the standard mass-loss rates shed approximately $10\,\msun$ after mass-transfer, ending their lives as WC-type stars. Meanwhile, primaries calculated with the new mass-loss recipe lose only about $5\,\msun$ after mass transfer and conclude their evolution as H-free WN stars, without transitioning to the WC phase. Such a huge difference in the mass drastically alters predictions for the formation mass of black holes in environments with different metallicities and hence, also has implications for the merger rates of compact objects. 

        A similar pattern emerges in the LMC models. Binary evolution models using the new mass-loss recipe fail to produce WC-type stars under the explored initial conditions. However, it remains plausible that WC stars could form when not using MLT++, with recalibrated mixing efficiencies, or in binaries with shorter orbital periods where mass transfer is more efficient in removing the envelope. Interestingly, the few known WC binaries in the LMC have orbital periods ranging between $\SIrange{2}{15}{d}$ \citep{bar2:01}, suggesting that systems with tighter initial configurations may favor WC star formation even under the new mass-loss regime. 
        Alternatively, as discussed in Sect.\,\ref{sec:WRmdot} it remains possible that we underestimate the mass-loss in the classical, hydrogen-poor WR stage as these are underrepresented in our curated sample. 

        At SMC metallicity, the paradigm shifts. Models using the standard mass-loss recipes evolve into H-free WN stars and spend most of their lives on the He-ZAMS. This contrasts with observations, as no H-free WN stars are detected at this metallicity. Conversely, the models calculated with our new mass-loss recipe lose significantly less mass, retaining surface hydrogen abundances around $X_\mathrm{H}\approx0.25$, consistent with observational data. These models also exhibit a broader temperature distribution, potentially accounting for the observed scatter in the temperatures of WR stars at SMC metallicity. This difference highlights the importance of adopting realistic mass-loss prescriptions in stellar evolution models, particularly in low-metallicity environments where mass loss plays a critical role in shaping the observable properties of massive stars. 

        We conclude that the outcomes of single and binary star evolutionary models strongly depend on the assumptions on the mass-loss rate on the OB and WR stages. The new, empiric, mass-loss rate recipe (Eq.~\eqref{eq:mdotfit}) pinned to the Eddington parameter readily supplied by stellar evolution models, eliminates the need to switch and interpolate between various free-choice theoretical prescriptions, as usually done in the literature.     
        
    \section{Summary}
    \label{sec:conclusions}

        In this work, we have compiled a sample of 183 massive stars with well-constrained stellar and wind properties from literature across the Galaxy (70 stars), LMC (61 stars), and SMC (52 stars). To ensure a consistent dataset capable of yielding robust conclusions, each star included in our sample passed several quality checks. The resulting sample spans initial masses from ${M_\mathrm{ini}\approx15\,\msun\text{\,--\,}150\,\msun}$ and temperatures in the range ${T\approx\SIrange{12}{100}{kK}}$, encompassing stars at various evolutionary phases. This makes our dataset well-suited for deriving general insights into stellar winds and mass-loss rates. Based on this data, we established that the mass-loss rates of hot stars can to first order be described as a function of the Eddington $\Gamma_{\!\text{e}}$ and metallicity, $Z$. Our newly derived mass-loss relation (see Eq.~\ref{eq:mdotfit}) has a root-mean-square dispersion of ${\Delta \log(\dot{M}/(\msunpyr)) = 0.43}$, corresponding to a factor of 2.7 in mass-loss rates.

        We compared our mass-loss scaling relation to mass-loss prescriptions typically used in stellar evolution calculations. 
        As many other works on individual targets or small samples of stars have reported, we also find that the typically applied mass-loss recipes overestimate the empirically derived mass-loss rates of hot massive stars by several factors. 
        
        We calculated stellar evolution models using two sets of mass-loss rates: one employing standard mass-loss recipes and another using our newly derived rates to explore how they impact stellar evolution and their final outcomes. The comparison highlights that the standard mass-loss rates overestimate mass loss even during the main sequence. This overestimation may contribute to the long-standing mass-discrepancy problem, where spectroscopic and evolutionary mass estimates diverge. The enhanced accuracy of our mass-loss rates also provides an opportunity to refine mixing efficiency parameters, which play a critical role in determining stellar structure and evolution.

        To bypass uncertainties associated with advanced evolutionary stages, we computed a small test set of binary evolutionary models using either mass-loss recipes. Interestingly, the models with our mass-loss rates did not produce WC-type stars within the explored parameter space. Whether this is due to the limited initial parameter selection, the calibration of mixing efficiencies, or another factor remains subject to future work. However, our models effectively explain the lack of H-free WN-type stars at low metallicity, where models with standard mass-loss rates fall short. The inclusion of our new empiric mass-loss recipe has profound implications for the feedback mechanisms, ionizing flux, and final evolutionary outcomes of massive stars in low-metallicity environments as in the young Universe.
                
    \begin{acknowledgements}
        We thank the referee for thoughtful comments that improved the manuscript. DP acknowledges financial support from the Deutsches Zentrum f\"ur Luft und Raumfahrt (DLR) grant FKZ 50OR2005 and the FWO junior postdoctoral fellowship No. 1256225N.
        AACS, MBP, RRL, and VR are supported by the Deutsche Forschungsgemeinschaft (DFG - German Research Foundation) in the form of an Emmy Noether Research Group -- Project-ID 445674056 (SA4064/1-1, PI Sander).
        JJ acknowledges funding from the Deutsche Forschungsgemeinschaft (DFG, German Research Foundation) Project-ID 496854903 (SA4064/2-1, PI Sander) and is a member of the International Max Planck Research School for Astronomy and Cosmic Physics at the University of Heidelberg (IMPRS-HD). JSV acknowledges support from STFC (Science and Technology Facilities Council) funding under grant number ST/V000233/1.
        The collaboration of coauthors was facilitated by support International Space Science Institute (ISSI) in Bern, through ISSI International Team project 512 (Multiwavelength View on Massive Stars in the Era of Multimessenger Astronomy, PI Oskinova).
    \end{acknowledgements}

 	\bibliographystyle{aa}                                                         
 	\bibliography{astro}

 \clearpage

   \begin{appendix}
   \onecolumn
 		\section{Tables}
 		\label{app:tables}
        {\small \vspace{-0.3cm}
        \begin{longtable}{lccccccccl}
 	          \caption{Stellar and wind parameters of OB stars with surface gravities below $\log(g/(\mathrm{cm\,s^{-2}}))<3.8$.}\label{tab:stellar_parameters_summary_supergiants} \\
 	        \hline \hline \rule{0cm}{2.2ex}%
            \rule{0cm}{2.2ex} Name & $T$            & $\log(L)$  & $\log(g)$                & $\varv\,\sin i$           & $M^\dagger$   & $X_H$     & $\log(\Gamma_{\!\text{e}})$  & $\log(\dot{M})$        & reference         \\
            \rule{0cm}{2.2ex}      & [$\mathrm{kK}$]& [$\lsun$]  & [$\mathrm{cm\,s^{-2}}$]  & [$\mathrm{km\,s^{-1}}$]   & [$\msun$]     & [mass fr.]&                   & [$\msunpyr$]           &                   \\
            \hline%
            \endfirsthead 
            
            \multicolumn{10}{l}{{{\normalsize \tablename\ \thetable{} continued.}}} \vspace{0.3cm}\\
 	        \hline \hline \rule{0cm}{2.2ex}%
            \rule{0cm}{2.2ex} Name & $T$            & $\log(L)$  & $\log(g)$                & $\varv\,\sin i$           & $M^\dagger$   & $X_H$     & $\log(\Gamma_{\!\text{e}})$  & $\log(\dot{M})$        & reference         \\
            \rule{0cm}{2.2ex}      & [$\mathrm{kK}$]& [$\lsun$]  & [$\mathrm{cm\,s^{-2}}$]  & [$\mathrm{km\,s^{-1}}$]   & [$\msun$]     & [mass fr.]&                   & [$\msunpyr$]           &                   \\
            \hline%
 			\endhead
    
     		\hline
     	    \multicolumn{10}{l}{{{
            \begin{minipage}{0.95\linewidth}
                \ignorespaces 
                 \rule{0cm}{2.8ex}$^\dagger$ Masses are calculated using $\log(g)$ corrected for the centrifugal force. $^{\dagger\dagger}$ Mass-loss rate corrected for a clumping factor of $D=10$.
            \end{minipage}}}}\\
            \endfoot
                
     		\hline
     	    \multicolumn{10}{l}{{{
            \begin{minipage}{0.95\linewidth}
                \ignorespaces 
                 \rule{0cm}{2.8ex}$^\dagger$ Masses are calculated using $\log(g)$ corrected for the centrifugal force. $^{\dagger\dagger}$ Mass-loss rate corrected for a clumping factor of $D=10$.
            \end{minipage}}}}\\
            \endlastfoot
            
            \multicolumn{10}{c}{\rule{0cm}{2.4ex}GAL}\\
            \hline \rule{0cm}{2.4ex}%
                Cyg\,OB2\,\#12      & 13.7         & 6.28        & 1.70                  & 38                      & 111.9         & 0.748     &  -0.34            & -5.52                  & \citet{cla1:12} \\
                CD-47\,4551        & 38.0           & 6.19       & 3.60                     & 50                        & 120.9         & 0.649     &  -0.49            & -5.54$^{\dagger\dagger}$& \citet{mar1:18} \\
                HD\,169582         & 37.0           & 6.10       & 3.50                     & 73                        & 86.1          & 0.414     &  -0.50            & -5.69$^{\dagger\dagger}$& \citet{mar1:18} \\
                $\zeta$\,Sco       & 17.2           & 6.05       & 2.00                     & 45                        & 53.3          & 0.748     &  -0.25            & -5.75                   & \citet{cla1:12} \\
                HD\,152408         & 31.8           & 5.98       & 3.00                     & 80                        & 39.0          & 0.270     &  -0.32            & -4.94                   & \citet{cro1:09} \\
                HD\,190429A        & 39.0           & 5.96       & 3.60                     & 150                       & 66.2          & 0.627     &  -0.46            & -5.96                   & \citet{bou1:12} \\
                HD\,16691          & 41.0           & 5.94       & 3.65                     & 135                       & 57.7          & 0.627     &  -0.42            & -5.52                   & \citet{bou1:12} \\
                HD\,15570          & 39.0           & 5.94       & 3.50                     & 40                        & 48.5          & 0.716     &  -0.32            & -5.66                   & \citet{bou1:12} \\
                $\epsilon$\,Ori    & 27.0           & 5.92       & 3.00                     & 70                        & 63.9          & 0.733     &  -0.46            & -6.07$^{\dagger\dagger}$& \citet{pub1:16} \\
                HD\,93843a         & 39.0           & 5.91       & 3.65                     & 90                        & 64.1          & 0.707     &  -0.48            & -5.85$^{\dagger\dagger}$& \citet{mar1:18} \\
                HD\,66811          & 40.0           & 5.91       & 3.60                     & 210                       & 55.7          & 0.611     &  -0.44            & -5.70                   & \citet{bou1:12} \\
                HD\,152408         & 30.5           & 5.90       & 3.00                     & 80                        & 38.3          & 0.490     &  -0.32            & -5.17                   & \citet{cro1:09} \\
                HD\,151804         & 29.0           & 5.90       & 3.00                     & 104                       & 47.6          & 0.430     &  -0.44            & -5.20                   & \citet{cro1:09} \\
                HD\,14947          & 37.0           & 5.83       & 3.50                     & 130                       & 48.1          & 0.677     &  -0.44            & -5.85                   & \citet{bou1:12} \\
                HD\,210839         & 36.0           & 5.80       & 3.50                     & 210                       & 53.0          & 0.677     &  -0.51            & -5.85                   & \citet{bou1:12} \\
                HD\,69464          & 36.0           & 5.78       & 3.50                     & 83                        & 46.9          & 0.707     &  -0.47            & -6.05$^{\dagger\dagger}$& \citet{mar1:18} \\
                HD\,163758         & 34.5           & 5.76       & 3.40                     & 94                        & 42.4          & 0.627     &  -0.47            & -5.80                   & \citet{bou1:12} \\
                HD\,148546	       & 31.0           & 5.70       & 3.20                     & 100                       & 35.7          & 0.414     &  -0.52            & -5.75$^{\dagger\dagger}$& \citet{mar1:18} \\
                HD\,192639         & 33.5           & 5.68       & 3.40                     & 90                        & 39.6          & 0.627     &  -0.52            & -5.92                   & \citet{bou1:12} \\
                HD\,152003         & 30.5           & 5.66       & 3.15                     & 77                        & 30.7          & 0.707     &  -0.41            & -5.92$^{\dagger\dagger}$& \citet{mar1:18} \\
                HD\,188209         & 29.8           & 5.65       & 3.20                     & 45                        & 36.7          & 0.707     &  -0.50            & -6.40                   & \citet{mar2:15} \\
                HD\,167264         & 28.0           & 5.65       & 3.10                     & 70                        & 37.9          & 0.707     &  -0.51            & -6.50                   & \citet{mar2:15} \\
                HD\,149404a        & 34.0           & 5.63       & 3.55                     & 67                        & 46.5          & 0.707     &  -0.62            & -6.03                   & \citet{rau2:16} \\
                HD\,75211          & 34.0           & 5.63       & 3.50                     & 145                       & 43.3          & 0.619     &  -0.61            & -6.64$^{\dagger\dagger}$& \citet{mar1:18} \\
                HD\,78344          & 30.0           & 5.60       & 3.15                     & 64                        & 28.7          & 0.414     &  -0.52            & -5.83$^{\dagger\dagger}$& \citet{mar1:18} \\
                HD\,152249         & 31.5           & 5.59       & 3.20                     & 65                        & 25.7          & 0.707     &  -0.40            & -6.06$^{\dagger\dagger}$& \citet{mar1:18} \\
                HD\,76968a         & 31.0           & 5.58       & 3.25                     & 55                        & 29.8          & 0.707     &  -0.48            & -6.11$^{\dagger\dagger}$& \citet{mar1:18} \\
                HD\,149404b        & 25.9           & 5.58       & 3.05                     & 272                       & 38.5          & 0.707     &  -0.70            & -6.48                   & \citet{rau2:16} \\
                HD\,190603         & 18.0           & 5.58       & 2.10                     & 49                        & 19.3          & 0.748     &  -0.28            & -5.96                   & \citet{cla1:12} \\
                HD\,168625         & 14.0           & 5.58       & 1.74                     & 60                        & 24.0          & 0.582     &  -0.42            & -6.38$^{\dagger\dagger}$& \citet{mah1:16} \\
                HD\,198478         & 16.0           & 5.57       & 2.15                     & 38                        & 33.1          & 0.716     &  -0.52            & -6.30                   & \citet{ber1:23} \\
                HD\,15629          & 41.0           & 5.56       & 3.75                     & 90                        & 29.8          & 0.751     &  -0.49            & -6.5                    & \citet{mar2:05} \\
                HD\,75222          & 30.0           & 5.56       & 3.15                     & 67                        & 25.7          & 0.707     &  -0.44            & -6.03$^{\dagger\dagger}$& \citet{mar1:18} \\
                CD-43\,4690        & 37.0           & 5.53       & 3.60                     & 91                        & 29.5          & 0.707     &  -0.53            & -6.41$^{\dagger\dagger}$& \citet{mar1:18} \\
                HD\,63005a         & 38.5           & 5.52       & 3.75                     & 63                        & 34.4          & 0.561     &  -0.64            & -6.79$^{\dagger\dagger}$& \citet{mar1:18} \\
                HD\,94963a         & 36.0           & 5.47       & 3.50                     & 82                        & 23.1          & 0.707     &  -0.47            & -6.32$^{\dagger\dagger}$& \citet{mar1:18} \\
                HD\,302505	       & 34.0           & 5.43       & 3.60                     & 43                        & 32.4          & 0.707     &  -0.67            & -6.76$^{\dagger\dagger}$& \citet{mar1:18} \\
                HD\,94370a         & 36.0           & 5.36       & 3.70                     & 185                       & 29.9          & 0.707     &  -0.70            & -6.30$^{\dagger\dagger}$& \citet{mar1:18} \\
                HD\,209975         & 30.5           & 5.35       & 3.35                     & 40                        & 23.7          & 0.707     &  -0.60            & -6.50                   & \citet{mar2:15} \\
                HD\,53138          & 16.0           & 5.14       & 2.15                     & 38                        & 12.4          & 0.716     &  -0.52            & -6.63                   & \citet{ber1:23} \\
                CD-44\,4865        & 30.0           & 5.26       & 3.45                     & 60                        & 26.0          & 0.707     &  -0.74            & -6.87$^{\dagger\dagger}$& \citet{mar1:18} \\
                HD\,46966          & 35.0           & 5.20       & 3.75                     & 50                        & 24.1          & 0.751     &  -0.75            & -7.50                   & \citet{mar1:12} \\
                HD\,69106          & 30.0           & 5.09       & 3.30                     & 310                       & 29.8          & 0.707     &  -0.77            & -7.35$^{\dagger\dagger}$& \citet{mar1:18} \\
                HD\,206165         & 18.0           & 5.07       & 2.45                     & 39                        & 13.1          & 0.716     &  -0.62            & -6.67                   & \citet{ber1:23} \\
                HD\,207198         & 32.5           & 5.05       & 3.50                     & 60                        & 13.1          & 0.707     &  -0.65            & -7.00                   & \citet{mar2:15} \\
                HD\,164353         & 15.0           & 4.05       & 2.50                     & 25                        & 2.9           & 0.716     &  -0.99            & -8.15                   & \citet{ber1:23} \\
            \hline%
            \multicolumn{10}{c}{\rule{0cm}{2.4ex}LMC}\\
            \hline \rule{0cm}{2.4ex}%
                R136b              & 37.0           & 6.35       & 3.40                     & 85                        & 123.2         & 0.700     &  -0.33            & -5.11$^{\dagger\dagger}$& \citet{bes2:20} \\
                N206-FS\,187       & 38.0           & 6.28       & 3.60                     & 116                       & 150.0         & 0.742     &  -0.47            & -5.59                   & \citet{ram2:18} \\
                Sk-68\,135         & 26.9           & 6.10       & 2.81                     & 45                        & 63.6          & 0.640     &  -0.30            & -5.70                   & Alkousa et al. (submitted) \\
                HDE\,269896        & 27.5           & 5.97       & 2.70                     & 70                        & 34.3          & 0.414     &  -0.23            & -5.12                   & \citet{eva1:04} \\
                Sk-67\,5           & 25.6           & 5.89       & 2.80                     & 65                        & 47.3          & 0.530     &  -0.41            & -6.05                   & Alkousa et al. (submitted) \\
                Sk-68\,52          & 26.0           & 5.87       & 2.85                     & 50                        & 47.2          & 0.700     &  -0.38            & -6.28                   & Alkousa et al. (submitted) \\
                N206-FS\,147       & 20.0           & 5.78       & 2.50                     & 40                        & 48.9          & 0.742     &  -0.48            & -6.13                   & \citet{ram1:18} \\
                HDE\,269050        & 25.5           & 5.76       & 2.70                     & 80                        & 29.0          & 0.414     &  -0.36            & -6.04                   & \citet{eva1:04} \\
                Sk-67\,2           & 18.8           & 5.76       & 2.30                     & 45                        & 38.1          & 0.640     &  -0.42            & -6.21                   & Alkousa et al. (submitted) \\
                Sk-66\,171         & 29.9           & 5.67       & 3.10                     & 75                        & 30.7          & 0.570     &  -0.43            & -6.07                   & Alkousa et al. (submitted) \\
                Sk-68\,155         & 29.0           & 5.64       & 3.05                     & 80                        & 29.0          & 0.570     &  -0.44            & -6.19                   & Alkousa et al. (submitted) \\
                Sk-69\,279         & 28.5           & 5.63       & 2.95                     & 40                        & 23.6          & 0.640     &  -0.34            & -5.70                   & Alkousa et al. (submitted) \\
                PGMW\,3053         & 34.7           & 5.60       & 3.50                     & 88                        & 36.0          & 0.742     &  -0.53            & -6.10                   & \citet{gom1:25} \\
                PGMW\,3061         & 42.0           & 5.60       & 3.70                     & 120                       & 26.9          & 0.742     &  -0.40            & -6.00                   & \citet{gom1:25} \\
                Sk-69\,52          & 18.8           & 5.60       & 2.30                     & 50                        & 26.6          & 0.570     &  -0.44            & -6.62                   & Alkousa et al. (submitted) \\
                Sk-67\,14          & 21.1           & 5.58       & 2.50                     & 50                        & 25.2          & 0.570     &  -0.44            & -6.33                   & Alkousa et al. (submitted) \\
                Sk-68\,8           & 14.1           & 5.57       & 1.80                     & 35                        & 24.7          & 0.600     &  -0.43            & -6.50                   & Alkousa et al. (submitted) \\
                Sk-69\,43          & 22.4           & 5.55       & 2.70                     & 50                        & 29.2          & 0.700     &  -0.50            & -6.49                   & Alkousa et al. (submitted) \\
                Sk-71\,41          & 29.2           & 5.53       & 3.10                     & 45                        & 24.1          & 0.700     &  -0.43            & -6.03                   & Alkousa et al. (submitted) \\
                Sk-68\,140         & 24.1           & 5.52       & 2.80                     & 50                        & 25.6          & 0.700     &  -0.47            & -6.46                   & Alkousa et al. (submitted) \\
                PGMW\,3100         & 38.0           & 5.50       & 3.70                     & 99                        & 31.5          & 0.742     &  -0.57            & -6.20                   & \citet{gom1:25} \\
                N206-FS\,134       & 30.0           & 5.50       & 3.40                     & 90                        & 40.7          & 0.742     &  -0.68            & -6.09                   & \citet{ram1:18} \\
                PGMW\,3168         & 33.3           & 5.30       & 3.50                     & 55                        & 21.0          & 0.742     &  -0.60            & -6.70                   & \citet{gom1:25} \\
                N206-FS\,58        & 25.0           & 5.05       & 3.00                     & 100                       & 12.6          & 0.742     &  -0.62            & -7.95                   & \citet{ram1:18} \\
            \hline%
            \multicolumn{10}{c}{\rule{0cm}{2.4ex}SMC}\\
            \hline \rule{0cm}{2.4ex}%
                AzV\,78            & 19.5           & 5.96       & 2.15                     & 26                        & 36.5          & 0.550     &  -0.23            & -5.82                   & \citet{ber1:24} \\
                AzV\,75            & 38.5           & 5.94       & 3.51                     & 120                       & 52.5          & 0.649     &  -0.38            & -6.30                   & \citet{bou1:21} \\
                AzV\,232           & 33.5           & 5.89       & 3.16                     & 75                        & 36.1          & 0.414     &  -0.33            & -5.92                   & \citet{bou1:21} \\
                AzV\,488           & 25.5           & 5.88       & 2.75                     & 55                        & 41.7          & 0.620     &  -0.34            & -6.07                   & \citet{ber1:24} \\
                AzV\,15            & 39.0           & 5.83       & 3.61                     & 120                       & 48.6          & 0.707     &  -0.44            & -6.46                   & \citet{bou1:21} \\
                AzV\,235           & 27.5           & 5.82       & 3.05                     & 39                        & 52.9          & 0.738     &  -0.48            & -6.17                   & \citet{ber1:24} \\
                Sk\,191            & 19.0           & 5.67       & 2.25                     & 57                        & 27.0          & 0.580     &  -0.38            & -6.02                   & \citet{ber1:24} \\
                AzV\,215           & 27.0           & 5.63       & 2.90                     & 86                        & 27.0          & 0.600     &  -0.41            & -6.44                   & \citet{ber1:24} \\
                AzV\,242           & 23.0           & 5.62       & 2.60                     & 40                        & 24.4          & 0.670     &  -0.36            & -6.39                   & \citet{ber1:24} \\
                AzV\,69            & 33.9           & 5.61       & 3.50                     & 70                        & 40.1          & 0.707     &  -0.57            & -6.51                   & \citet{bou1:21} \\
                AzV\,362           & 14.5           & 5.58       & 1.80                     & 14                        & 22.1          & 0.680     &  -0.35            & -6.92                   & \citet{ber1:24} \\
                AzV\,83            & 32.8           & 5.54       & 3.26                     & 80                        & 22.2          & 0.414     &  -0.47            & -6.14                   & \citet{bou1:21} \\
                AzV\,327           & 30.0           & 5.54       & 3.12                     & 95                        & 23.9          & 0.561     &  -0.46            & -7.37                   & \citet{bou1:21} \\
                AzV\,95            & 38.0           & 5.46       & 3.70                     & 55                        & 28.3          & 0.707     &  -0.57            & -7.40                   & \citet{bou1:21} \\
                AzV\,104           & 26.5           & 5.45       & 3.05                     & 73                        & 26.7          & 0.600     &  -0.58            & -6.96                   & \citet{ber1:24} \\
                AzV\,18            & 19.0           & 5.45       & 2.35                     & 41                        & 20.1          & 0.738     &  -0.43            & -6.66                   & \citet{ber1:24} \\
                AzV\,47            & 35.0           & 5.44       & 3.75                     & 60                        & 42.2          & 0.707     &  -0.76            & -8.18                   & \citet{bou1:21} \\
                AzV\,266           & 23.0           & 5.44       & 2.65                     & 44                        & 18.2          & 0.500     &  -0.46            & -6.92                   & \citet{ber1:24} \\
                AzV\,264           & 21.5           & 5.43       & 2.50                     & 44                        & 16.6          & 0.550     &  -0.41            & -6.82                   & \citet{ber1:24} \\
                AzV\,77            & 37.5           & 5.40       & 3.74                     & 150                       & 29.7          & 0.561     &  -0.69            & -7.88                   & \citet{bou1:21} \\
                SMCSGS-FS\,310     & 32.0           & 5.40       & 3.60                     & 100                       & 39.6          & 0.738     &  -0.77            & -7.90                   & \citet{ram1:19} \\
                AzV\,210           & 19.5           & 5.39       & 2.45                     & 25                        & 19.6          & 0.738     &  -0.48            & -6.66                   & \citet{ber1:24} \\
                AzV\,187           & 16.5           & 5.33       & 2.20                     & 40                        & 19.0          & 0.738     &  -0.52            & -7.22                   & \citet{ber1:24} \\
                AzV\,472           & 19.5           & 5.29       & 2.50                     & 32                        & 17.5          & 0.738     &  -0.53            & -7.52                   & \citet{ber1:24} \\
                SMCSGS-FS\,310     & 30.0           & 5.27       & 3.20                     & 300                       & 22.3          & 0.738     &  -0.65            & -6.60                   & \citet{ram1:19} \\
                AzV\,22            & 14.0           & 5.27       & 1.90                     & 42                        & 16.3          & 0.738     &  -0.52            & -7.15                   & \citet{ber1:24} \\
                AzV\,439           & 31.0           & 5.16       & 3.54                     & 260                       & 26.7          & 0.678     &  -0.86            & -7.70                   & \citet{bou1:21} \\
                AzV\,307           & 30.0           & 5.15       & 3.50                     & 60                        & 22.7          & 0.649     &  -0.80            & -8.82                   & \citet{bou1:21} \\
                AzV\,170           & 30.5           & 5.14       & 3.51                     & 70                        & 21.3          & 0.649     &  -0.78            & -8.82                   & \citet{bou1:21} \\
                AzV\,43            & 28.5           & 5.13       & 3.37                     & 200                       & 22.6          & 0.561     &  -0.84            & -8.00                   & \citet{bou1:21} \\
                AzV\,234           & 16.5           & 4.97       & 2.40                     & 14                        & 12.9          & 0.738     &  -0.71            & -7.74                   & \citet{ber1:24} \\
                MPG\,012           & 31.0           & 4.93       & 3.65                     & 60                        & 16.9          & 0.737     &  -0.87            & -9.30                   & \citet{bou1:13} \\
                Sk\,179            & 14.5           & 4.81       & 2.30                     & 83                        & 13.3          & 0.738     &  -0.89            & -8.00                   & \citet{ber1:24} \\
                AzV\,343           & 12.5           & 4.72       & 2.05                     & 45                        & 10.3          & 0.738     &  -0.87            & -8.00                   & \citet{ber1:24} \\
        \end{longtable}
        }

        \begin{table}[b]
            \centering
            \caption{Stellar and wind parameters of partially stripped stars.}
            \small
            \begin{tabular}{lccccccccl}\hline \hline \rule{0cm}{2.8ex}%
                \rule{0cm}{2.2ex} Name & $T$            & $\log(L)$  & $\log(g)$                & $\varv\,\sin i$           & $M^\dagger$   & $X_H$     & $\log(\Gamma_{\!\text{e}})$  & $\log(\dot{M})$        & reference         \\
                \rule{0cm}{2.2ex}      & [$\mathrm{kK}$]& [$\lsun$]  & [$\mathrm{cm\,s^{-2}}$]  & [$\mathrm{km\,s^{-1}}$]   & [$\msun$]     & [mass fr.]&                   & [$\msunpyr$]           &                   \\
                \hline%
                \multicolumn{10}{c}{\rule{0cm}{2.4ex}LMC}\\
                \hline \rule{0cm}{2.4ex}%
                    Sk-71\,35          & 20.0           & 4.93       & 2.60                     & 85                        & 9.5           & 0.70      &  -0.63            & -7.00                   & \citet{ram1:24} \\
                \hline%
                \multicolumn{10}{c}{\rule{0cm}{2.4ex}SMC}\\
                \hline \rule{0cm}{2.4ex}%
                    SSN\,7             & 43.6           & 5.75       & 3.60                     & 135                       & 26.4          & 0.600     &  -0.28            & -5.40                   & \citet{ric1:23} \\
                    AzV\,476           & 42.0           & 5.65       & 3.70                     & 140                       & 30.5          & 0.737     &  -0.41            & -6.10                   & \citet{pau2:22} \\
                    2dFS\,2553         & 42.0           & 4.91       & 3.80                     & 80                        & 6.9           & 0.600     &  -0.54            & -7.30                   & \citet{ram1:23} \\
                    2dFS\,163          & 37.0           & 4.75       & 3.50                     & 60                        & 4.0           & 0.330     &  -0.54            & -6.90                   & \citet{ram1:23} \\
                    SMCSGS-FS\,69      & 24.0           & 4.70       & 2.65                     & 50                        & 2.9           & 0.600     &  -0.37            & -6.20                   & \citet{ram1:23} \\
                \hline
            \end{tabular}
            \rule{0cm}{2.8ex}%
            \begin{minipage}{0.95\linewidth}
                \ignorespaces 
                 $^\dagger$ Masses are calculated using $\log(g)$ corrected for the centrifugal force. 
            \end{minipage}
            \label{tab:stellar_parameters_summary_partially_stripped}
        \end{table}

        \begin{table}[tbp]
            \centering
            \caption{Stellar and wind parameters of WR binaries.}
            \small
            \begin{tabular}{lccccccl}\hline \hline \rule{0cm}{2.8ex}%
                \rule{0cm}{2.2ex} Name & $T$            & $\log(L)$  & $M_\mathrm{orb}^\dagger$& $X_H$     & $\log(\Gamma_{\!\text{e}})$  & $\log(\dot{M})$        & reference         \\
                \rule{0cm}{2.2ex}      & [$\mathrm{kK}$]& [$\lsun$]  & [$\msun$]               & [mass fr.]&                   & [$\msunpyr$]           &                   \\
                \hline%
                \multicolumn{8}{c}{\rule{0cm}{2.4ex}GAL}\\
                \hline \rule{0cm}{2.4ex}%
                    WR\,102a           & 34.1           & 6.445      & 82                      & 0.229      &  -0.19            & -4.43                 & \citet{loh1:18} \\
                    WR\,102b           & 33.8           & 6.305      & 60                      & 0.502      &  -0.11            & -4.84                 & \citet{loh1:18} \\
                    WR\,22             & 44.7           & 6.30       & 71.7                    & 0.400      &  -0.22            & -4.85                 & \citet{rau1:96,gra1:08} \\
                    WR\,133            & 67.0           & 5.35       & 7.9                     & 0.000      &  -0.36            & -5.05                 & \citet{ric1:21} \\
                    WR\,137            & 60.0           & 5.22       & 5.3$^{a}$               & 0.000      &  -0.32            & -4.65                 & \citet{lev2:05,ric1:16} \\
                \hline%
                \multicolumn{8}{c}{\rule{0cm}{2.4ex}LMC}\\
                \hline \rule{0cm}{2.4ex}%
                    R144a                  & 45.0           & 6.44       & 74                      & 0.35      &  -0.11            & -4.38                   & \citet{she1:21} \\
                    Mk\,34a                & 53.0           & 6.43       & 147                     & 0.65      &  -0.22            & -4.88                   & \citet{teh1:19} \\
                    R144b                  & 40.0           & 6.39       & 69                      & 0.40      &  -0.12            & -4.34                   & \citet{she1:21} \\
                    Mk\,34b                & 53.0           & 6.37       & 136                     & 0.65      &  -0.35            & -4.72                   & \citet{teh1:19} \\
                    BAT99\,119a            & 45.0           & 6.35       & 53                      & 0.40      &  -0.04            & -4.40                   & \citet{she1:19} \\
                    BAT99\,119b            & 40.0           & 6.33       & 54                      & 0.50      &  -0.04            & -4.20                   & \citet{she1:19} \\
                    BAT99\,113             & 45.0           & 6.14       & 53                      & 0.70      &  -0.17            & -5.50                   & \citet{she1:19} \\
                    BAT99\,103             & 46.0           & 6.09       & 26                      & 0.20      &  -0.06            & -5.00                   & \citet{she1:19} \\
                    BAT99\,107             & 32.0           & 6.09       & 63                      & 0.70      &  -0.29            & -5.20                   & \citet{she1:19} \\
                    BAT99\,129             & 89.0           & 5.81       & 16                      & 0.10      &  -0.37            & -5.40                   & \citet{she1:19} \\
                    BAT99\,077             & 41.0           & 5.79       & 22                      & 0.70      &  -0.13            & -5.20                   & \citet{she1:19} \\
                    BAT99\,043             & 71.0           & 5.69       & 14                      & 0.30      &  -0.16            & -4.90                   & \citet{she1:19} \\
                    BAT99\,049             & 96.0           & 5.35       & 11                      & 0.30      &  -0.39            & -5.60                   & \citet{she1:19} \\
                \hline%
                \multicolumn{8}{c}{\rule{0cm}{2.4ex}SMC}\\
                \hline \rule{0cm}{2.4ex}%
                    AB\,3a                 & 77.0           & 5.93       & 20                      & 0.25     &  -0.12            & -5.30                   & \citet{she1:16} \\
                    AB\,5a                 & 43.0           & 6.35       & 61                      & 0.25     &  -0.15            & -4.50                   & \citet{she1:16} \\
                    AB\,5b                 & 43.0           & 6.25       & 66                      & 0.25     &  -0.26            & -4.50                   & \citet{she1:16} \\
                    AB\,7a                 & 98.0           & 6.10       & 23                      & 0.15     &  -0.02            & -5.00                   & \citet{she1:16} \\
                    AB\,8a                 & 115.0          & 6.15       & 19                      & 0.00     &   0.05            & -4.80                   & \citet{she1:16} \\
                    AB\,6a                 & 80.0           & 5.87       & 16                      & 0.25     &  -0.05            & -5.20                   & \citet{she1:18} \\
                \hline
            \end{tabular}
            \rule{0cm}{2.8ex}%
            \begin{minipage}{0.95\linewidth}
                \ignorespaces 
                 $^\dagger$ Note that when the inclination is unknown, it is typically estimated by matching the secondary's spectroscopic mass to the orbital one. For more details, we refer to the individual papers. $^{a}$ We assumed an inclination of $i=59.7^\circ$ to match the spectroscopic mass (corrected for the centrifugal force) of the OB star with the orbital mass.
            \end{minipage}
            \label{tab:stellar_parameters_summary_WR}
        \end{table}

        \begin{table}[b]
            \centering
            \caption{Stellar and wind parameters of [WR] stars.}
            \small
            \begin{tabular}{lccccccl}\hline \hline \rule{0cm}{2.8ex}%
                \rule{0cm}{2.2ex} Name & $T$            & $\log(L)$  & $M^\dagger$  & $X_H$     & $\log(\Gamma_{\!\text{e}})$  & $\log(\dot{M})$        & reference         \\
                \rule{0cm}{2.2ex}      & [$\mathrm{kK}$]& [$\lsun$]  & [$\msun$]    & [mass fr.]&                   & [$\msunpyr$]           &                   \\
                \hline%
                \multicolumn{8}{c}{\rule{0cm}{2.4ex}GAL}\\
                \hline \rule{0cm}{2.4ex}%
                    $[$S71d$]$3        & 166.0          & 3.91       & 0.6           & 0.0      &  -0.68            & -6.65                  & \citet{tod1:15}   \\
                    NGC\,6905          & 139.0          & 3.90       & 0.6           & 0.0      &  -0.69            & -6.96                  & \citet{gom1:22}   \\
                    PC\,22             & 130.0          & 3.78       & 0.6           & 0.0      &  -0.81            & -7.36                  & \citet{sab1:22}   \\
                    NGC\,6369          & 124.0          & 3.78       & 0.6           & 0.0      &  -0.81            & -6.78                  & \citet{tod1:15}   \\
                    Abell\,48          & 70.0           & 3.77       & 0.6           & 0.1      &  -0.78            & -6.40                  & \citet{tod1:13}   \\
                    PB\,8              & 52.0           & 3.77       & 0.6           & 0.4      &  -0.67            & -7.10                  & \citet{tod1:10}   \\
                    NGC\,2867          & 157.0          & 3.68       & 0.6           & 0.0      &  -1.23            & -7.20                  & \citet{tod1:15}   \\
                    NGC\,5189          & 157.0          & 3.60       & 0.6           & 0.0      &  -0.99            & -7.34                  & \citet{tod1:15}   \\
                    IC\,4663           & 140.0          & 3.60       & 0.6           & 0.02     &  -0.98            & -7.70                  & \citet{mis1:12}   \\
                    PB\,6              & 157.0          & 3.56       & 0.6           & 0.0      &  -1.03            & -7.25                  & \citet{tod1:15}   \\
                    NGC\,2371          & 130.0          & 3.45       & 0.6           & 0.0      &  -1.14            & -7.75                  & \citet{gom1:20}   \\
                    Hen\,2-55          & 126.0          & 3.43       & 0.6           & 0.0      &  -1.16            & -7.54                  & \citet{tod1:15}   \\
                \hline
            \end{tabular}
            \rule{0cm}{2.8ex}%
            \begin{minipage}{0.95\linewidth}
                \ignorespaces 
                 $^\dagger$ For [WR] stars no spectroscopic mass can be derived. Model calculations predict that these objects have masses between ${\sim0.4\,\msun\text{\,--\,}0.8\,\msun}$ \citep{tyl1:03}. For simplicity the average mass of $0.6\,\msun$ is assumed. Note that these stars are not included in the fitting routine due to this uncertainty.
            \end{minipage}
            \label{tab:stellar_parameters_summary_bracket_WR}
        \end{table}
        
        \begin{table}[tbp]
            \centering
            \caption{Stellar and wind parameters of OB stars with surface gravities above $\log(g/(\mathrm{cm\,s^{-2}}))\geq3.8$.}
            \small
            \begin{tabular}{lccccccccl}\hline \hline \rule{0cm}{2.8ex}%
                \rule{0cm}{2.2ex} Name & $T$            & $\log(L)$  & $\log(g)$                & $\varv\,\sin i$           & $M^\dagger$   & $X_H$     & $\log(\Gamma_{\!\text{e}})$  & $\log(\dot{M})$        & reference         \\
                \rule{0cm}{2.2ex}      & [$\mathrm{kK}$]& [$\lsun$]  & [$\mathrm{cm\,s^{-2}}$]  & [$\mathrm{km\,s^{-1}}$]   & [$\msun$]     & [mass fr.]&                   & [$\msunpyr$]           &                   \\
                \hline%
                \multicolumn{10}{c}{\rule{0cm}{2.4ex}GAL}\\
                \hline \rule{0cm}{2.4ex}%
                    HD\,93250          & 44.0           & 6.12       & 4.00                     & 110                       & 144.1         & 0.751     &  -0.60            & -6.25                   & \citet{mar2:05}  \\
                    HD\,64568a         & 48.0           & 5.80       & 4.00                     & 55                        & 48.5          & 0.707     &  -0.46            & -6.43$^{\dagger\dagger}$& \citet{mar1:18} \\
                    HD\,93204a         & 40.5           & 5.70       & 3.90                     & 105                       & 60.9          & 0.707     &  -0.67            & -6.40$^{\dagger\dagger}$& \citet{mar1:18} \\
                    HD\,46150          & 42.0           & 5.65       & 4.01                     & 100                       & 60.3          & 0.751     &  -0.71            & -7.30                   & \citet{mar1:12} \\
                    HD\,46223          & 43.0           & 5.60       & 4.01                     & 100                       & 49.0          & 0.751     &  -0.67            & -7.17                   & \citet{mar1:12} \\
                    HD\,93204          & 40.0           & 5.51       & 4.00                     & 130                       & 52.4          & 0.751     &  -0.78            & -6.75                   & \citet{mar2:05}  \\
                    CPD-59\,2600a      & 40.0           & 5.40       & 4.00                     & 120                       & 40.3          & 0.766     &  -0.77            & -6.46$^{\dagger\dagger}$& \citet{mar1:18} \\
                    HD\,91824a         & 39.0           & 5.37       & 3.90                     & 47                        & 32.7          & 0.707     &  -0.73            & -7.32$^{\dagger\dagger}$& \citet{mar1:18} \\
                    HD\,93222          & 38.0           & 5.36       & 3.90                     & 52                        & 35.2          & 0.707     &  -0.77            & -6.71$^{\dagger\dagger}$& \citet{mar1:18} \\
                    HD\,91572a         & 38.5           & 5.35       & 3.90                     & 49                        & 32.7          & 0.707     &  -0.75            & -6.70$^{\dagger\dagger}$& \citet{mar1:18} \\
                    HD\,42088          & 38.0           & 5.23       & 4.00                     & 60                        & 33.3          & 0.751     &  -0.86            & -8.00                   & \citet{mar2:05}  \\
                    HD\,93146          & 37.0           & 5.22       & 4.00                     & 80                        & 36.3          & 0.751     &  -0.91            & -7.25                   & \citet{mar2:05}  \\
                    HD\,46485          & 36.0           & 5.05       & 3.85                     & 300                       & 23.3          & 0.751     &  -0.90            & -7.30                   & \citet{mar1:12} \\
                    HD\,93028          & 34.0           & 5.05       & 4.00                     & 50                        & 34.2          & 0.751     &  -1.05            & -9.00                   & \citet{mar2:05}  \\
                    HD\,97848	       & 36.5           & 5.03       & 3.90                     & 42                        & 19.6          & 0.707     &  -0.84            & -7.22$^{\dagger\dagger}$& \citet{mar1:18} \\
                    CPD-58\,2620a      & 38.5           & 4.99       & 3.95                     & 39                        & 16.0          & 0.707     &  -0.80            & -7.50$^{\dagger\dagger}$& \citet{mar1:18} \\
                    HD\,92504          & 35.0           & 4.99       & 3.85                     & 155                       & 19.7          & 0.707     &  -0.89            & -7.63$^{\dagger\dagger}$& \citet{mar1:18} \\
                    HD\,66788          & 34.0           & 4.96       & 4.00                     & 55                        & 27.9          & 0.738     &  -1.06            & -8.92                  & \citet{mar3:09}  \\
                    HD\,46202	       & 34.0           & 4.88       & 4.00                     & 15                        & 22.8          & 0.707     &  -1.06            & -7.69$^{\dagger\dagger}$& \citet{mar1:18} \\
                    $\zeta$\,Oph       & 32.0           & 4.86       & 3.60                     & 400                       & 18.5          & 0.738     &  -0.98            & -8.80                  & \citet{mar3:09}  \\
                    HD\,46056          & 34.5           & 4.85       & 3.89                     & 330                       & 20.0          & 0.751     &  -1.03            & -8.00                  & \citet{mar1:12} \\
                    HD\,152590         & 36.0           & 4.79       & 4.10                     & 66                        & 18.9          & 0.751     &  -1.06            & -7.78                  & \citet{mar2:05}  \\
                    HD\,34078          & 33.0           & 4.77       & 4.05                     & 40                        & 22.7          & 0.751     &  -1.15            & -9.50                  & \citet{mar2:05}  \\
                    HD\,326329         & 31.0           & 4.74       & 3.90                     & 80                        & 19.5          & 0.738     &  -1.12            & -9.22                  & \citet{mar3:09}  \\
                    HD\,38666          & 33.0           & 4.66       & 4.00                     & 111                       & 16.1          & 0.751     &  -1.11            & -9.50                  & \citet{mar2:05}  \\
                    $\tau$\,Sco        & 30.7           & 4.30       & 3.97                     & 5                         & 8.5           & 0.738     &  -1.20            & -8.60                  & \citet{osk1:11}  \\
                \hline%
                \multicolumn{10}{c}{\rule{0cm}{2.4ex}LMC}\\
                \hline \rule{0cm}{2.4ex}%
                    H36                & 48.0           & 6.29       & 4.00                     & 125                       & 150.8         & 0.75      &  -0.46            & -5.28$^{\dagger\dagger}$& \citet{bes2:20} \\
                    R136a6             & 53.0           & 6.27       & 4.10                     & 160                       & 122.8         & 0.74      &  -0.39            & -5.65$^{\dagger\dagger}$& \citet{bes2:20} \\
                    H31                & 48.0           & 6.01       & 4.00                     & 130                       & 79.6          & 0.74      &  -0.46            & -6.28$^{\dagger\dagger}$& \citet{bes2:20} \\
                    H58                & 50.0           & 5.94       & 4.10                     & 150                       & 72.7          & 0.74      &  -0.49            & -7.12$^{\dagger\dagger}$& \citet{bes2:20} \\
                    H40                & 45.0           & 5.88       & 3.90                     & 150                       & 61.4          & 0.74      &  -0.48            & -6.58$^{\dagger\dagger}$& \citet{bes2:20} \\
                    H55                & 47.0           & 5.76       & 3.90                     & 130                       & 39.0          & 0.76      &  -0.40            & -6.42$^{\dagger\dagger}$& \citet{bes2:20} \\
                    H62                & 49.0           & 5.75       & 4.00                     & 170                       & 41.2          & 0.74      &  -0.44            & -6.31$^{\dagger\dagger}$& \citet{bes2:20} \\
                    H35                & 44.0           & 5.74       & 4.00                     & 180                       & 61.7          & 0.76      &  -0.62            & -6.38$^{\dagger\dagger}$& \citet{bes2:20} \\
                    H65                & 42.0           & 5.74       & 3.90                     & 160                       & 58.8          & 0.76      &  -0.60            & -6.67$^{\dagger\dagger}$& \citet{bes2:20} \\
                    H68                & 43.0           & 5.73       & 4.00                     & 210                       & 66.8          & 0.76      &  -0.66            & -7.39$^{\dagger\dagger}$& \citet{bes2:20} \\
                    H50                & 42.0           & 5.71       & 3.80                     & 200                       & 45.1          & 0.76      &  -0.51            & -6.67$^{\dagger\dagger}$& \citet{bes2:20} \\
                    H64                & 40.0           & 5.69       & 3.90                     & 180                       & 64.2          & 0.76      &  -0.69            & -6.88$^{\dagger\dagger}$& \citet{bes2:20} \\
                    H52                & 44.0           & 5.67       & 4.00                     & 180                       & 52.7          & 0.76      &  -0.62            & -6.42$^{\dagger\dagger}$& \citet{bes2:20} \\
                    H66                & 46.0           & 5.64       & 4.10                     & 115                       & 50.6          & 0.76      &  -0.63            & -6.15$^{\dagger\dagger}$& \citet{bes2:20} \\
                    H78                & 48.0           & 5.60       & 4.20                     & 105                       & 48.8          & 0.76      &  -0.66            & -6.50$^{\dagger\dagger}$& \citet{bes2:20} \\
                    H94                & 48.0           & 5.52       & 4.20                     & 170                       & 40.1          & 0.76      &  -0.67            & -7.00$^{\dagger\dagger}$& \citet{bes2:20} \\
                    PGMW\,3058         & 40.0           & 5.40       & 3.90                     & 66                        & 31.9          & 0.742     &  -0.68            & -6.50                   & \citet{gom1:25} \\
                    H92                & 39.0           & 5.26       & 4.00                     & 150                       & 33.0          & 0.76      &  -0.83            & -7.56$^{\dagger\dagger}$& \citet{bes2:20} \\
                    H80                & 35.0           & 5.15       & 3.80                     & 155                       & 25.4          & 0.76      &  -0.82            & -7.66$^{\dagger\dagger}$& \citet{bes2:20} \\
                    H114               & 44.0           & 5.25       & 4.20                     & 100                       & 30.9          & 0.74      &  -0.81            & -7.33$^{\dagger\dagger}$& \citet{bes2:20} \\
                    H108               & 43.0           & 5.04       & 4.20                     & 260                       & 22.8          & 0.76      &  -0.89            & -7.81$^{\dagger\dagger}$& \citet{bes2:20} \\
                    H123               & 41.0           & 5.01       & 4.10                     & 120                       & 19.0          & 0.74      &  -0.84            & -7.98$^{\dagger\dagger}$& \citet{bes2:20} \\
                    H134               & 36.0           & 4.81       & 4.00                     & 105                       & 16.0          & 0.76      &  -0.96            & -8.17$^{\dagger\dagger}$& \citet{bes2:20} \\
                \hline%
                \multicolumn{10}{c}{\rule{0cm}{2.4ex}SMC}\\
                \hline \rule{0cm}{2.4ex}%
                    MPG\,355           & 51.7           & 6.04       & 4.00                     & 120                       & 63.3          & 0.737     &  -0.34            & -6.74                   & \citet{bou1:13} \\
                    SMCSGS-FS\,231	   & 46.0           & 5.61       & 4.10                     & 100                       & 47.0          & 0.738     &  -0.64            & -7.50                   & \citet{ram1:19} \\
                    AzV\,243           & 39.6           & 5.59       & 3.90                     & 60                        & 51.3          & 0.737     &  -0.69            & -7.10                   & \citet{bou1:13} \\
                    AzV\,388           & 43.1           & 5.54       & 4.01                     & 150                       & 43.0          & 0.737     &  -0.67            & -7.00                   & \citet{bou1:13} \\
                    AzV\,177           & 44.5           & 5.43       & 4.03                     & 220                       & 32.1          & 0.737     &  -0.65            & -6.85                   & \citet{bou1:13} \\
                    AzV\,14a           & 42.8           & 5.41       & 4.00                     & 90                        & 31.5          & 0.737     &  -0.66            & -7.70                   & \citet{pau1:23} \\
                    AzV\,14b           & 41.8           & 5.38       & 4.00                     & 90                        & 32.3          & 0.737     &  -0.70            & -7.70                   & \citet{pau1:23} \\
                    MPG\,396           & 37.0           & 5.30       & 4.00                     & 196                       & 45.4          & 0.737     &  -0.93            & -8.70                   & \citet{ric1:22} \\
                    MPG\,113           & 39.6           & 5.15       & 4.00                     & 35                        & 23.4          & 0.737     &  -0.79            & -8.52                   & \citet{bou1:13} \\
                    AzV\,461           & 37.1           & 5.00       & 4.05                     & 200                       & 25.7          & 0.737     &  -0.98            & -9.00                   & \citet{bou1:13} \\
                    NGC346-31          & 37.2           & 4.95       & 4.00                     & 25                        & 18.9          & 0.737     &  -0.90            & -9.22                   & \citet{bou1:13} \\
                    MPG\,356           & 38.2           & 4.88       & 4.10                     & 20                        & 18.2          & 0.737     &  -0.95            & -8.46                   & \citet{bou1:13} \\
                \hline
            \end{tabular}
            \rule{0cm}{2.8ex}%
            \begin{minipage}{0.95\linewidth}
                \ignorespaces 
                 $^\dagger$ Masses are calculated using $\log(g)$ corrected for the centrifugal force.
            \end{minipage}
            \label{tab:stellar_parameters_summary_main_sequence}
        \end{table}

        \twocolumn
        \section{MESA modeling}

        \label{app:mesa}

            For our stellar evolution calculations, we utilized the Modules for Experiments in Stellar Astrophysics (MESA) code, version 24.08.1 \citep{pax1:11,pax1:13,pax1:15,pax1:18,pax1:19,jer1:23}. The physical setup follows the framework established by \citet{bro1:11}. Convection is modeled using the Ledoux criterion and the standard mixing length theory \citep{boe1:58}, adopting a mixing length parameter of $\alpha_\mathrm{mlt}=1.5$. Semiconvective mixing in layers that are stable according to the Schwarzschild criterion but not the Ledoux criterion is included with an efficiency parameter of $\alpha_\mathrm{sc}=1$. For overshooting, we apply step overshooting for H and He convective cores, allowing the cores to extend by $0.335\,H_\mathrm{P}$, where $H_\mathrm{P}$ is the pressure scale height at the convective boundary \citep{bro1:11}. Additionally, thermohaline mixing is incorporated with an efficiency of $\alpha_\mathrm{th}=1$. Rotational mixing is treated as a diffusive process, accounting for dynamical and secular shear instabilities, the Goldreich-Schubert-Fricke instability, and Eddington-Sweet circulations \citep{heg1:00}. Consistent with \citet{bro1:11}, we adopt rotational mixing efficiency factors of $f_c=1/30$ and $f_\mu=0.1$. 

            The chemical abundances of H, He, C, N, O, Mg, Si, and Fe are tailored for the different galaxies following \citet{bro1:11}, with the used values listed in Table~\ref{tab:chem}. For elements not explicitly specified, abundances are scaled to the solar values from \citet{asp1:05} according to the metallicity of the galaxy. To reduce computation time all models are only evolved until core helium depletion. For the most massive stars, convergence issues were encountered during core He-burning. To address this, we applied the MLT++ prescription, as detailed in Sect. 7.2 of \citet{pax1:13}, for stars with helium core masses exceeding $M>12\,\msun$.

            Within the binary model calculations, mass transfer during the main sequence is modeled using the ``contact'' scheme implemented in MESA \citep{mar1:16}. For subsequent evolutionary stages, we employed the updated mass transfer scheme from \citet{mar1:21}, which accounts for outflows from the outer Lagrangian points.

            \begin{table}[ht]
                \centering
                \caption{Chemical abundances used in the stellar evolution models in mass fractions.}
                \small
                \begin{tabular}{lcc}\hline \hline \rule{0cm}{2.8ex}%
                    \rule{0cm}{2.2ex}   Element     & abundance (mass-fr.)           & Reference             \\
                    \hline
                    \multicolumn{3}{c}{\rule{0cm}{2.4ex}GAL}\\
                    \hline \rule{0cm}{2.4ex}%
                                        H           & 0.726         & $1-Y-Z$               \\
                                        He          & 0.265         & Interpol.$^{(a)}$     \\
                                        C           &$\num{1.16e-3}$& \cite{hun1:07}        \\
                                        N           &$\num{4.40e-4}$& \cite{hun1:08,hun1:09}\\
                                        O           &$\num{4.08e-3}$& \cite{hun1:08,hun1:09}\\
                                        Mg          &$\num{3.60e-4}$& \cite{hun1:07}        \\
                                        Si          &$\num{5.17e-4}$& \cite{hun1:07}        \\
                                        Fe          &$\num{1.01e-3}$& \cite{ven1:95}        \\
                    \hline
                    \multicolumn{3}{c}{\rule{0cm}{2.4ex}LMC}\\
                    \hline \rule{0cm}{2.4ex}%
                                        H           & 0.738         & $1-Y-Z$            \\
                                        He          & 0.257         & Interpol.$^{(a)}$  \\
                                        C           &$\num{4.94e-4}$& \cite{kur1:98}     \\
                                        N           &$\num{8.14e-5}$& \cite{kur1:98}     \\
                                        O           &$\num{2.62e-3}$& \cite{kur1:98}     \\
                                        Mg          &$\num{1.97e-4}$&\cite{hun1:07,hun1:09}\\
                                                    &               &\cite{tru1:07}      \\
                                        Si          &$\num{3.24e-4}$&\cite{hun1:07,hun1:09}\\
                                                    &               &\cite{tru1:07}      \\
                                        Fe          &$\num{4.59e-4}$&\cite{fer1:06}      \\
                    \hline
                    \multicolumn{3}{c}{\rule{0cm}{2.4ex}SMC}\\
                    \hline \rule{0cm}{2.4ex}%
                                        H           &  0.746        & $1-Y-Z$               \\
                                        He          & Interpol.$^{(a)}$     \\
                                        C           &$\num{2.08e-4}$& \cite{kur1:98}        \\
                                        N           &$\num{3.27e-5}$& \cite{kur1:98}        \\
                                        O           &$\num{1.13e-3}$& \cite{kur1:98}        \\
                                        Mg          &$\num{9.31e-5}$& \cite{hun1:07,hun1:09}\\
                                                    &               & \cite{tru1:07}\\
                                        Si          &$\num{1.30e-4}$& \cite{hun1:07,hun1:09}\\
                                                    &               & \cite{tru1:07}\\
                                        Fe          &$\num{2.49e-4}$& \cite{ven1:99}\\
                    \hline
                \end{tabular}
                \rule{0cm}{3.2ex}%
                \begin{minipage}{0.95\linewidth}
                    \ignorespaces 
                     $^{(a)}$ Interpolated between the primordial He mass-fraction of $Y=0.2477$ \citep{pei1:07} and solar value $Y=0.28$ from \citep{gre1:96}. Note that in \citep{bro1:11} the GAL composition was chosen to match the abundances of the Galactic FLAMES survey.
                \end{minipage}
                \label{tab:chem}
            \end{table}
        
    \end{appendix}

\end{document}